\documentclass[aps,physrev,twocolumn,groupedaddress]{revtex4-2}
\usepackage{siunitx}% Correct display of units
\usepackage{graphicx}% Include figure files
\usepackage{subcaption}% For subfigure environment
\usepackage{dcolumn}% Align table columns on decimal point
\usepackage{bm}% bold math
\usepackage{wasysym} % diameter symbol

\usepackage{amsmath} %align
\begin{document}

\preprint{PRAB/CTR-Compression}

%Title of paper
\title{First demonstration of coherent radiation imaging for bunch-by-bunch longitudinal compression monitoring}

% repeat the \author .. \affiliation  etc. as needed
% \email, \thanks, \homepage, \altaffiliation all apply to the current
% author. Explanatory text should go in the []'s, actual e-mail
% address or url should go in the {}'s for \email and \homepage.
% Please use the appropriate macro foreach each type of information

% \affiliation command applies to all authors since the last
% \affiliation command. The \affiliation command should follow the
% other information
% \affiliation can be followed by \email, \homepage, \thanks as well.
\author{Joseph Wolfenden}
\email[corresponding author: ]{joseph.wolfenden@cockcroft.ac.uk}
\author{Ana Guisao-Betancur}
\author{Carsten Welsch}
\affiliation{Department of Physics, University of Liverpool, Liverpool, United Kingdom}
\affiliation{Cockcroft Institute, Daresbury, United Kingdom}

\author{Billy Kyle}
\altaffiliation{ISIS Neutron and Muon Source, STFC Rutherford Appleton Laboratory, Didcot, United Kingdom}
\affiliation{Department of Physics, University of Manchester, Manchester, United Kingdom}
\affiliation{Cockcroft Institute, Daresbury, United Kingdom}

\author{Thomas Pacey}
\affiliation{ASTeC, STFC Daresbury Laboratory, Daresbury, United Kingdom}
\affiliation{Cockcroft Institute, Daresbury, United Kingdom}

\author{Erik Mansten}
\author{Sara Thorin}
\author{Mathias Brandin}
\altaffiliation{European Spallation Source, SE-221 00 Lund, Sweden}
\affiliation{MAX IV Laboratory, Lund University, SE-224 84 Lund, Sweden}

%Collaboration name if desired (requires use of superscriptaddress
%option in \documentclass). \noaffiliation is required (may also be
%used with the \author command).
%\collaboration can be followed by \email, \homepage, \thanks as well.
%\collaboration{}
%\noaffiliation

\date{\today}

\begin{abstract}
Longitudinal bunch profile monitoring is a crucial diagnostic requirement in most accelerator facilities. This is particularly true in modern free-electron lasers and novel acceleration schemes, where bunch lengths are often $<$\SI{100}{\femto\second} and standard instrumentation is invasive or lacks the required resolution. This paper proposes a new monitoring method in this challenging parameter space. Initial proof of principle results for relative compression monitoring via broadband imaging of coherent THz radiation are presented. The technique can utilize more conventional intensity monitoring or novel spatial distribution variation. Both techniques have been demonstrated using both invasive and non-invasive coherent radiation sources. These results pave the way for a future non-invasive longitudinal bunch profile monitor.

\end{abstract}
% insert suggested keywords - APS authors don't need to do this
%\keywords{}

%\maketitle must follow title, authors, abstract, and keywords
\maketitle
% body of paper here - Use proper section commands
% References should be done using the \cite, \ref, and \label commands
\section{Introduction\label{sec:introduction}}
The longitudinal charge profile of a particle bunch is a key parameter in the optimization and operation of particle accelerators. This has become particularly important in modern free-electron lasers (FELs), where the longitudinal phase space (LPS) has a significant impact on lasing parameters and performance~\cite{Hanuka2021}, and in novel acceleration schemes, where the witness bunch length must be matched to the plasma wavelength for optimal acceleration~\cite{Adli2018}. The bunch lengths of interest in these applications are typically $<$\SI{100}{\femto \second}, which poses a considerable challenge for longitudinal diagnostics.

The gold standard for this type of measurement is the transverse deflecting cavity (TDC)~\cite{Burt2012}. This uses transverse RF modes within a dedicated cavity to produce a time-varying transverse kick to an electron bunch passing through the cavity. This impulse correlates the longitudinal axis of the bunch with a transverse axis, which can then be imaged as a standard transverse profile. A dispersive element, such as a dipole, is often integrated into this system to produce a full LPS measurement. TDCs can provide longitudinal charge profile resolutions on the $\sim$\SI{1}{\femto\second} scale. This measurement is perfectly suited to the vast majority of applications in FELs and novel acceleration, but it offers no pre-application monitoring capability, as it is a completely destructive measurement process. For bunch-by-bunch monitoring in these applications, TDC measurements can only be taken after the beam has been used, prior to the beam dump. Back-propagation to pre-application is then required, but this provides many application-specific challenges~\cite{Schrder2020, Christie2020}.
 
Monitoring has subtle differences to direct measurement. In direct measurements, the goal is to achieve the most detail at the highest resolution, and the survival of the bunch after this process is not a concern. In contrast, the main requirement for monitoring is non-invasive operation. The pursuit of this objective often leads to a compromise in other areas. For example, full profile reconstruction but over multiple bunches, or bunch-to-bunch operation at a reduced resolution or approximation of the bunch profile.

Rather than direct beam measurements, such as the TDC, most longitudinal charge profile monitoring techniques rely upon radiative processes, most of which fall under the category of polarization radiation~\cite{Karlovets2009}. Here, the electric field of the electron bunch is manipulated to produce electromagnetic radiation. It is this radiation that is then studied, rather than the bunch itself. If the radiation can be produced non-invasively, then a monitoring system is possible. For longitudinal measurements, it is the coherent spectrum of this radiation that is interrogated; for $<$\SI{100}{\femto \second}, the bandwidth of interest is in the THz. These measurements fall broadly into two areas: spectroscopy and interferometry. In both cases, the goal is to reconstruct the longitudinal form factor of the particle bunch. This is the well-known Fourier transform of the longitudinal bunch profile and provides a description of the frequency content of a particle bunch; the shorter the bunch, the higher the frequency content. In spectroscopic techniques, this form factor is directly sampled and reconstructed~\cite{Schmidt2018}. In interferometric techniques, an interferogram is produced that can then be used to infer the bunch form factor~\cite{Frohlich2005}.

Spectroscopy can mostly be conducted on a bunch-by-bunch basis but over limited bandwidths, meaning that extrapolations are required for sections of the bunch form factor that are not directly sampled. There are exceptions to this, but the complexity of the required system has led to limited uptake across the community. The main limitation in interferometric systems is the multi-shot operation; an interferogram requires many bunches to produce a single scan. This can also require additional correction factors, for example, modeled or measured frequency-dependent attenuation in the system. Again, there are bunch-by-bunch exceptions to this generalization~\cite{Thangaraj2012}, but these are only applicable to ultra-short bunches in the $\sim$\SI{1}{\femto\second} range. The limitations of existing methods demonstrate the requirement for novel techniques to monitor the longitudinal profile of high-brightness beams in a bunch-by-bunch manner.

This paper presents the development progress of such a system. It is the first demonstration of imaging broadband coherent polarization radiation to monitor and optimize longitudinal bunch compression. This is a simple and extremely flexible setup that is equally applicable to bunch lengths ranging from $>$\SI{100}{\femto\second} to below \SI{1}{\femto\second}. The work presented here is based upon a type of coherent polarization radiation known as coherent transition radiation (CTR). For the vast majority of cases, CTR cannot be considered non-invasive. However, it is simple to integrate and operate, which has expedited the development of the technique. Once the methodology has been established a non-invasive radiation source will replace the CTR source, namely coherent synchrotron radiation (CSR). A brief overview of established CTR theory and its application to the simulation of broadband imaging will be presented, showing how longitudinal information can be extracted from the images produced. Then, the initial experimental verification of these simulations is presented in two different setups. In each case, proof-of-concept (PoC) has been achieved, including first results from the implementation a non-invasive CSR radiation source. Finally, improvements to the system and the next stages of development will be discussed, including the changes required for non-invasive operation.
\section{Broadband Image Simulation\label{sec:simulation}}
\subsection{Coherent Transition Radiation}
Transition radiation (TR) is a broadband radiation source that is emitted as a charged particle passes through an interface between two materials of different dielectric properties; for particle accelerator applications, this is usually a vacuum and a metallic foil or mirror. In the relativistic limit, which is assumed from hereafter, either polarization charge analysis (PCA)~\cite{Karlovets2009} or pseudo-photon theory~\cite{Ter-Mikaelian1972} can be applied. A characteristic which must be considered here, which is usually ignored in more standard diagnostics (for example, optical TR-based profile monitors), is the thickness of the radiating interface. As this is a broadband source, the skin depth of the radiator will vary across the bandwidth of interest. Care must be taken to ensure that the thickness of this radiator exceeds the longest wavelength of interest in the TR spectrum, otherwise the target itself will act as a high-pass filter upon the radiated TR. This can have practical implications, such as in the suitability of standard TR foils or the thickness of deposited layers on mirrors.

\begin{figure}
\includegraphics[width=0.75\linewidth]{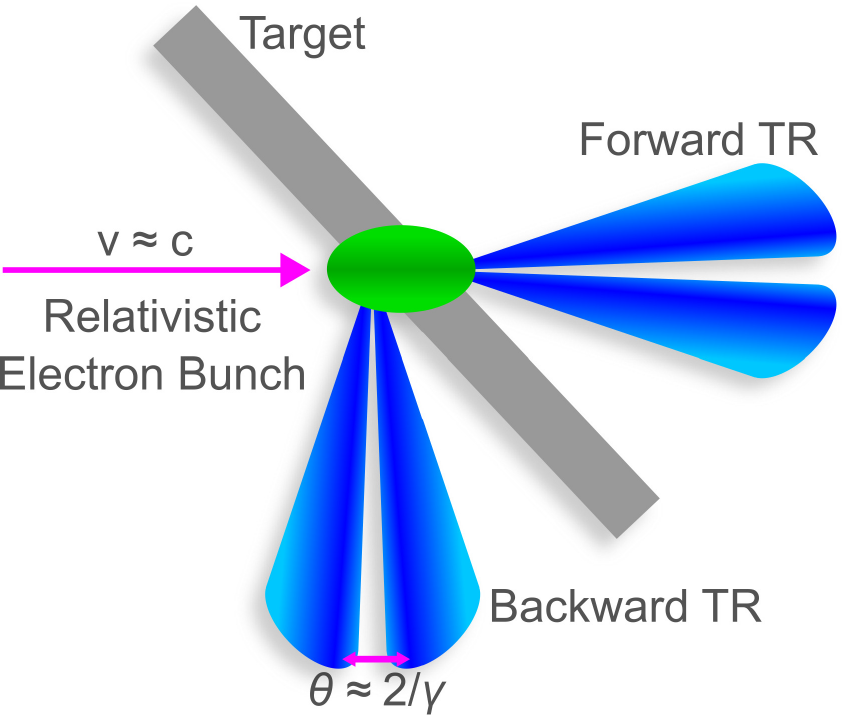}%
\caption{Schematic of the TR generation mechanism and propagation characteristics, where $v$ is the particle bunch velocity, $c$ is the speed of light, $\theta$ is the angle subtended by the peaks of the TR distribution, and $\gamma$ is the relativistic Lorentz factor for the particle bunch energy.\label{fig:tr_theory}}
\end{figure}

Radiation is emitted in a characteristic cone whose angular width is dependent upon the energy of the particle. This peaks at $\pm 1/\gamma$ in the far-field ($\gamma^2\lambda >> 1$), but can be significantly smaller in the near-field~\cite{Verzilov2000}. Radiation is emitted in both the forward and backward directions, as the particle enters and then exits the target. For diagnostic purposes, the target is often rotated relative to the direction of beam propagation, as the backward TR follows the path of specular reflection. Forward TR follows the direction of beam propagation regardless of the target orientation. This is illustrated in Fig. \ref{fig:tr_theory}.

To simulate the full broadband CTR image, the electric field definition of the source field are no different from any other TR source~\cite{Wolfenden2019, Karlovets2008, Kube2008}. What must be considered is the coherence of the specific frequency bandwidth. For a broadband TR source the effects of coherence vary across the large spectrum of emitted radiation. Longer wavelengths, which are more comparable to the spatial extent of the source particle bunch, will exhibit more coherence than shorter wavelengths. This is a notable effect due to the natural variation of the TR spatial distribution with frequency. Higher frequencies within the broadband TR spectrum produce narrower single particle spatial distributions~\cite{Kube2008, Wolfenden2019, Potylitsyn2020}, essentially a TR-based point spread function (PSF), which are therefore more intense due to the increased energy density. The total integral of these different spectral components will therefore vary depending on how they are modulated by the effect of the frequency-dependent coherence. Hence, the impact of coherence on the measured TR will vary depending on the properties of the source bunch. This variation forms the basis of this diagnostic technique and is encapsulated in the bunch form factor.

\begin{figure}
\includegraphics[width=1.0\linewidth]{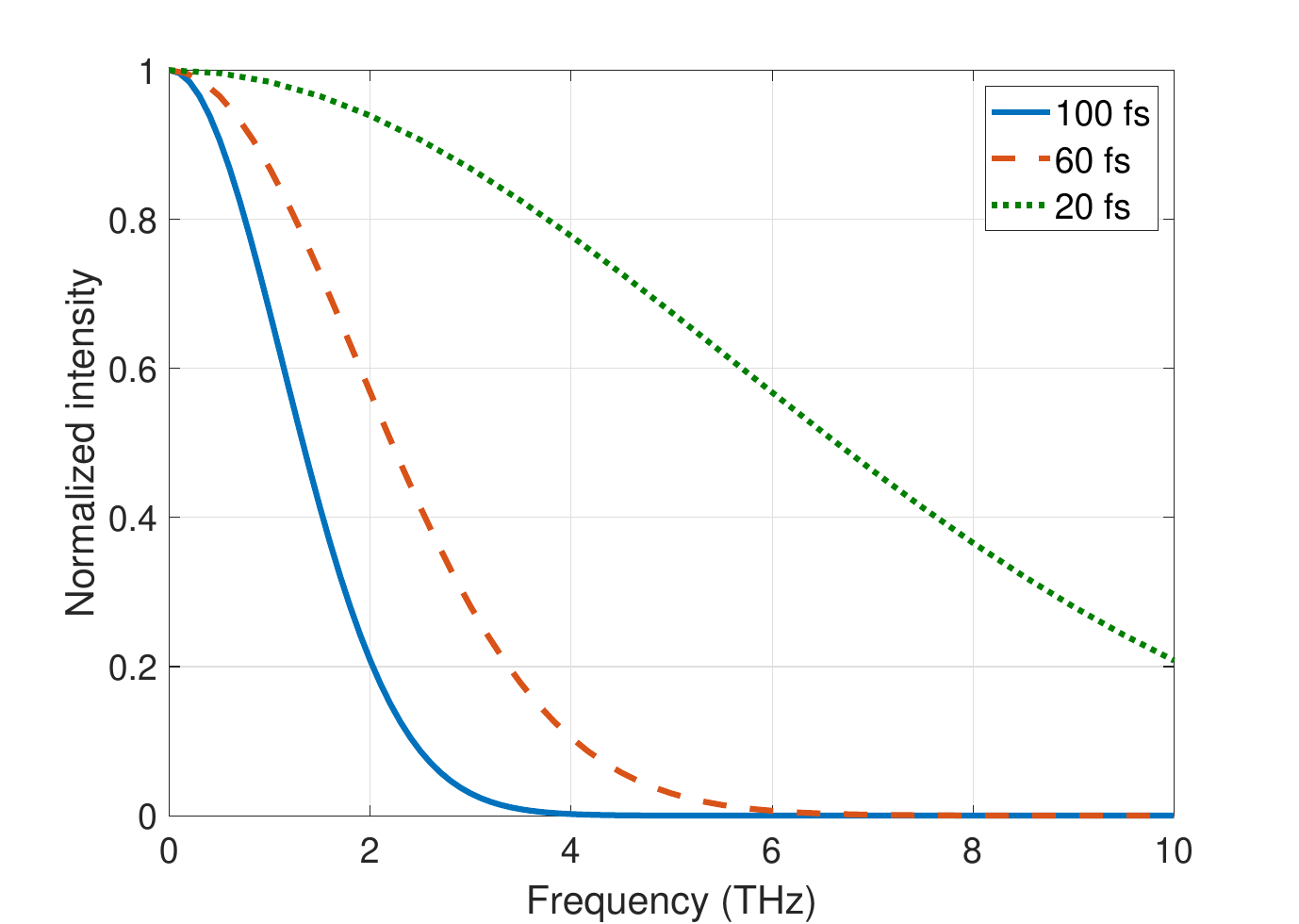}%
\caption{Demonstration of how the frequency content of a Gaussian bunch varies with bunch length.\label{fig:FF}}
\end{figure}

The form factor is defined as the Fourier transform of the spatial distribution of the source bunch charge~\cite{Castellano1999}. This can be split into transverse and longitudinal components, each dependent on the transverse and longitudinal distributions of the bunch, respectively. These values vary between $0$ and $1$ and serve to modulate the spectral intensity of the TR source. The formulation of the form factor, $F$, can be seen in Equ.~\ref{eq:FF}:
\begin{align}
\label{eq:FF}
    F(\omega) &= F_T(\omega)F_L(\omega) \nonumber \\
    &=\Bigg| \int{d^2r'e^{-ik\mathbf{r'}}S_T(\mathbf{r'})} \Bigg|^2 \Bigg| \int{dz'e^{-ikz'}S_L(z')} \Bigg|^2
\end{align}

where $\omega$ is the TR frequency and $S_T$ and $S_L$ are the spatial projections of the bunch distribution transversely and longitudinally, respectively. In most applications, the transverse distribution of the TR electric field can be considered much larger than the spatial extent of the particle distribution~\cite{Castellano1999}, leading to $F_T(\omega)\sim1$ across the bandwidth of interest for CTR, and the total form factor being defined entirely by the longitudinal distribution of the particle bunch. This assumption has been made throughout this work and was tested during experimental measurements. For situations where this assumption does not hold, Equ.~\ref{eq:FF} can be used to integrate the contributions from the transverse distribution~\cite{Potylitsyn2020}. Form factors for three Gaussian bunches are presented in Fig.~\ref{fig:FF}, which demonstrate how the frequency content of a bunch increases as the bunch length decreases. 

\subsection{Coherent Radiation Imaging}
The next step is to image this broadband TR source. In principle, this is no different than a standard point-to-point imaging system. The main characteristic which needs attention is the transmission response of the imaging optics. In broadband imaging, transmission can vary greatly across a bandwidth of interest, adding additional confounding components to the modulation of the spectral intensity of the TR. As discussed previously and as shown in Fig.~\ref{fig:FF}, there is a fixed bandwidth over which the distribution of the form factor fluctuates greatly; this is the region which will produce the greatest change in the CTR image as the bunch length is varied. The imaging system only needs to focus on this region and therefore this bandwidth is used to select materials for the imaging system. To accurately simulate a CTR image, this transmission curve of the imaging system needs to be well quantified. Other contributions to the modulation must also be taken into account; for example, the absorption of certain bands within air, or the quantum efficiency of the detector used.

The formulation of a TR source is well documented~\cite{Wolfenden2019, Karlovets2008, Kube2008}, but the literature usually focuses on optical TR (OTR). In order to account for broadband source, coherence must be accounted for. When studying coherent wavelengths, the imaging system is usually significantly within the near-field, or even pre-wave zone, of the TR source~\cite{Verzilov2000}. For angular imaging this may prove an issue, however, for spatial point-to-point imaging the resulting spatial distribution is largely unaffected assuming a sufficiently large limiting aperture is used, such that any vignetting is minimized~\cite{Xiang2007}. The TR electric field in the image plane of a spatial imaging system of a bunch of charged particles can therefore be represented as a coherent summation, i.e. superposition, of the image plane electric fields of the individual particles within the bunch~\cite{Loos2008}, as shown in Equ.~\ref{eq:TR_source_sum}.
\begin{equation}
\label{eq:TR_source_sum}
    E^i_{B}(\mathbf{r_i}) = \sum^N_{j=1}e^{-ikz_j}E^i_{P}(\mathbf{r_i - r_j}),
\end{equation}

where $E^i_B$ is the image plane electric field of a bunch of $N$ particles, $E^i_P$ is the image plane electric field of a single particle, and ($r_{j}$, $z_{j}$) is the transverse and longitudinal positions of the $j^{th}$ particle within a bunch, respectively. For large $N$, by taking the square of this field, Equ.~\ref{eq:TR_source_sum} can be refactored as an integral to calculate the spatial image plane spectral intensity at a given frequency,
\begin{align}
\label{eq:incoh_coh_intensity}
\frac{d^2U_B}{d\omega dr}\sim|E^i_B|^2 &= N\int d^2r'dz \rho(\mathbf{r'},z)|E^i_P(\mathbf{r_i}-\mathbf{r'})|^2 \nonumber \\
&+ N^2 \Bigg|\int d^2r'dz \rho(\mathbf{r'},z)E^i_P(\mathbf{r_i}-\mathbf{r'})\Bigg|^2,
\end{align}

where $\rho(\mathbf{r'},z)$ is the 3D bunch charge distribution. The term dependent on $N$ is the incoherent contribution to the final image, whilst the term dependent on $N^2$ is the coherent contribution to the final image. The coherent contribution completely dominates the image distribution due to the $N^2$ pre-factor. Therefore, the incoherent contribution can be disregarded~\cite{Loos2008} and, by combining Equ.~\ref{eq:FF} and ~\ref{eq:incoh_coh_intensity}, the image plane field intensity can be defined as:
\begin{equation}
    \label{eq:coh_intensity}
    \frac{dU_B(r_i)}{dr} = N^2\int_{\Delta\omega}{ |F_L(\omega)|^2\frac{d^2U_P(r_i)}{d\omega dr}d\omega},
\end{equation}

where the integral is taken over the bandwidth of interest, as defined by the bunch form factor, e.g. Fig.~\ref{fig:FF}. Equation~\ref{eq:coh_intensity} shows that the CTR bunch image distribution is actually a summation of spectral image distributions, modulated by the bunch form factor. When coupled with the natural variation of TR with frequency, that is, higher frequencies produce narrower and hence more intense field intensities~\cite{Kube2008, Wolfenden2019, Potylitsyn2020}, it follows that variations in the longitudinal bunch profile will produce variations in the bunch form factor, which will in turn produce variations in both shape and intensity of a broadband CTR image. The range of applicability of this method is defined purely by the bandwidth of the imaging system, which is itself defined by the materials used. Therefore, the same technique could be used to image both sub-fs and ps beams, but the materials used must allow for this shift in dominant frequencies within the bandwidth of interest.

\begin{figure}
     \centering
     \begin{subfigure}[b]{0.5\textwidth}
         \centering
         \includegraphics[width=\textwidth]{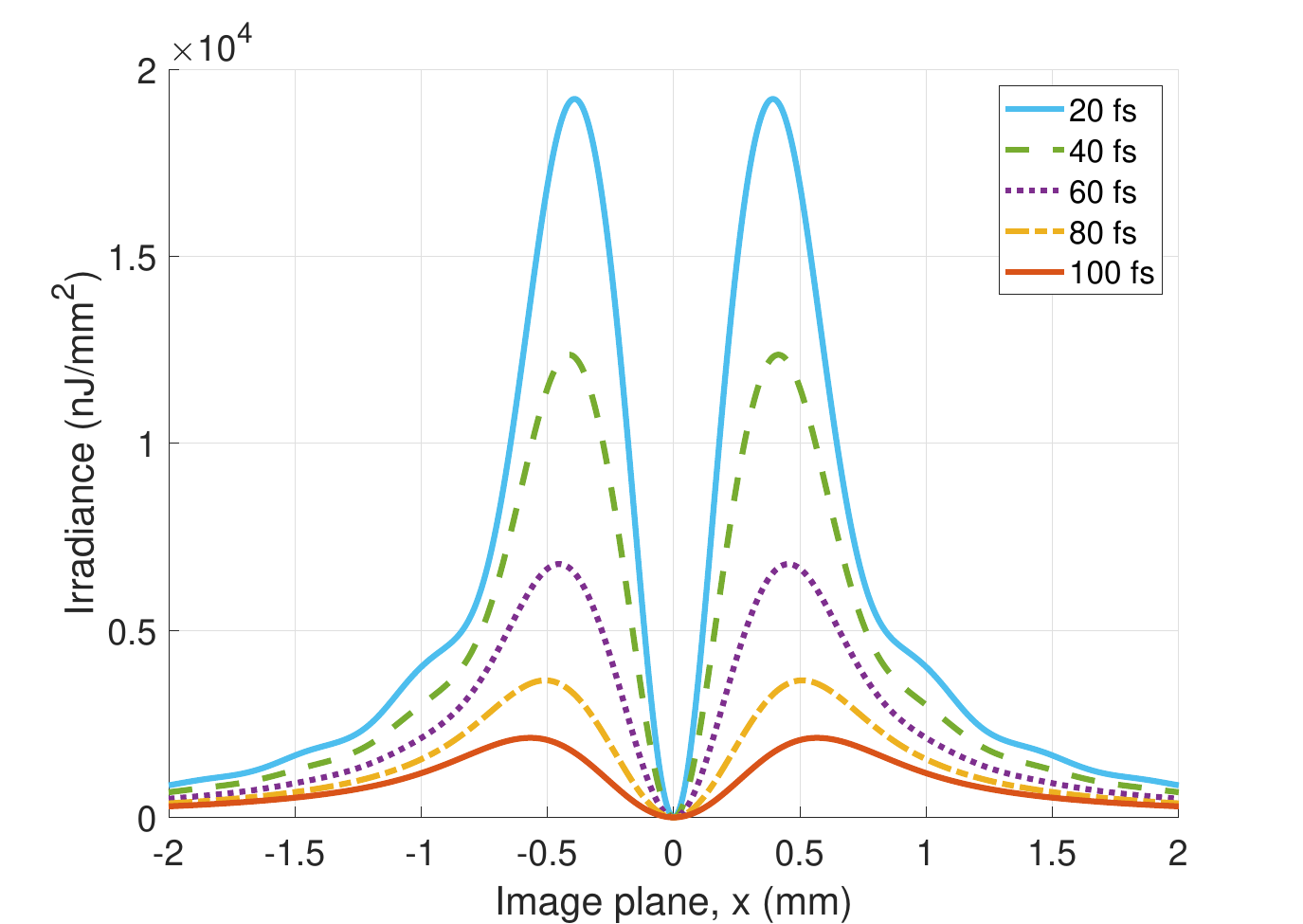}
         \caption{}
         \label{fig:tpx_simulation_abs}
     \end{subfigure}
     \vfill
     \begin{subfigure}[b]{0.5\textwidth}
         \centering
         \includegraphics[width=\textwidth]{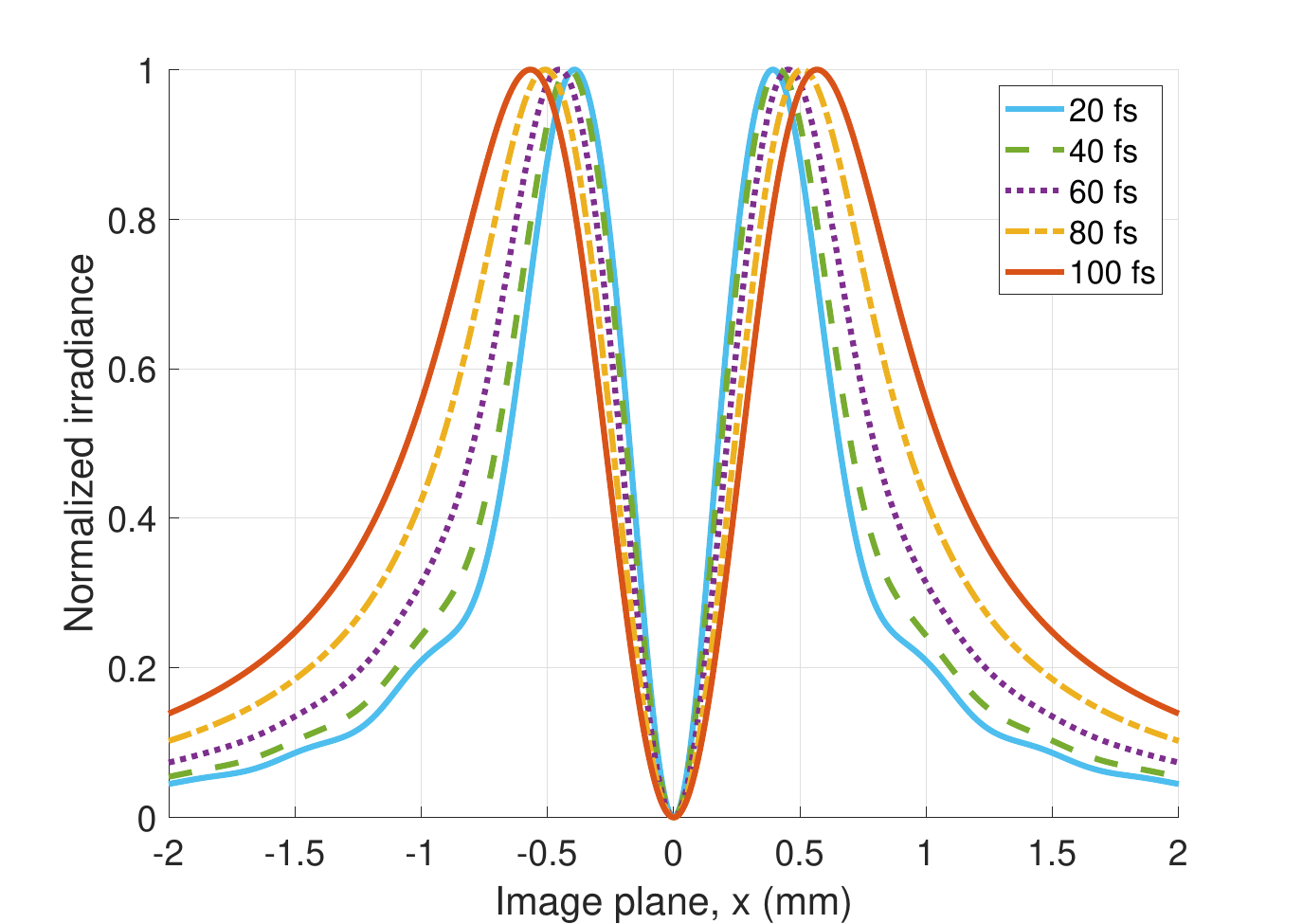}
         \caption{}
         \label{fig:tpx_simulation_norm}
     \end{subfigure}
        \caption{1D cross-sections of the absolute (a) and normalized (b) spatial distribution of a CTR image from a TPX lens 2f-2f imaging system for a range of Gaussian longitudinal bunch lengths.}
        \label{fig:tpx_simulation}
\end{figure}

This final point is nontrivial and requires some prior understanding of the bunch parameter space. To appropriately simulate the response of the different materials and the different optical components, the optical system simulation tool Ansys Zemax OpticStudio~\cite{Zemax} is used. This can propagate a custom electric field through a series of user-defined surfaces. As a demonstration of the above theory, the well-defined TR source distribution has been propagated through a single polymethylpentene (TPX)~\cite{Tydex} plano-convex lens (f = \SI{150}{\mm}, \diameter = \SI{50}{\mm}) imaging system in a 2f-2f configuration, where f is the focal length of the lens, using a bandwidth of \SI{0.1}{\THz} $-$ \SI{10}{\THz}. TPX was chosen for this example as it is a common material used in THz systems which has a high transmission across this bandwidth. Figure~\ref{fig:tpx_simulation} shows several 1D cross-sections of simulated 2D CTR image distributions for a range of Gaussian longitudinal bunch lengths, from \SI{100}{\femto\second} down to \SI{20}{\femto\second}. Figure~\ref{fig:tpx_simulation_abs} shows how the absolute intensity of the CTR image distribution varies with bunch length. This alone could function as a relative compression monitor but, due to the $N^2$ factor in Equ.~\ref{eq:coh_intensity}, the bunch charge would need to be monitored in parallel, to ensure fluctuations in intensity were due to bunch length variation and not bunch charge drift or noise. Instead, Fig.~\ref{fig:tpx_simulation_norm} presents the normalized distribution of the same CTR simulations. It is now clear that the shape of the distribution is also dependent upon the longitudinal bunch profile, which has the benefit of not being dependent upon the bunch charge.

\begin{figure}
\includegraphics[width=1.0\linewidth]{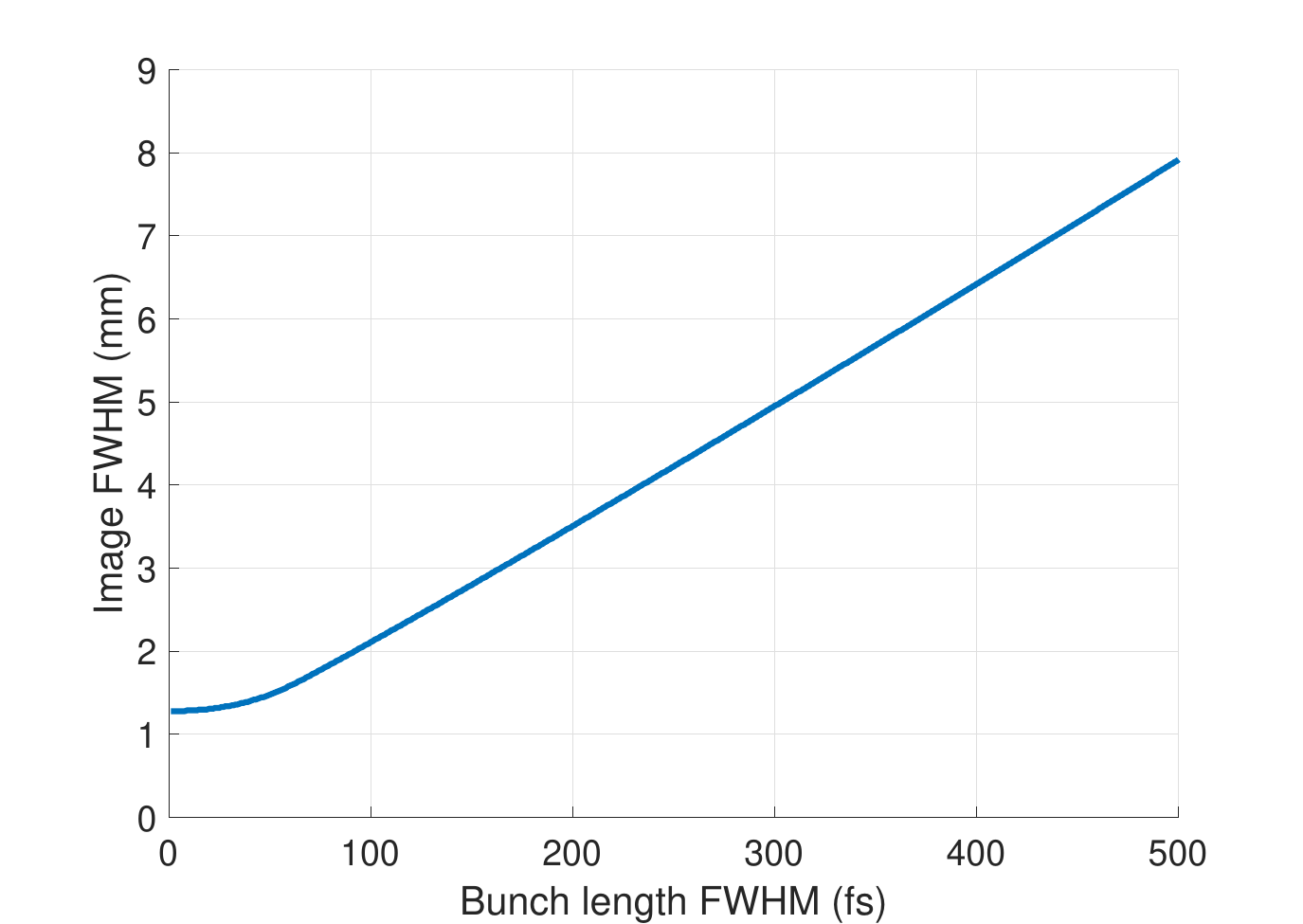}%
\caption{Demonstration of FWHM distance variation with bunch length in CTR image distributions for a TPX lens system.\label{fig:tpx_simulation_fwhm}}
\end{figure}

Figure~\ref{fig:tpx_simulation_fwhm} shows how, as an example metric, the FWHM of the distribution varies with bunch length. There is a clear linear region for bunch lengths $>$\SI{100}{\femto\second} that would function as a direct RMS measurement of bunch length. Important to note here is the region $<$\SI{100}{\femto\second} where the curve plateaus. This is the region that corresponds to the upper limit of the bandwidth that has been simulated, i.e. as the bunch length decreases, the frequency content increases, and the bandwidth of \SI{0.1}{\THz} $-$ \SI{10}{\THz} is no longer sufficient. This is evident in Fig.~\ref{fig:FF}, where the \SI{20}{\femto\second} bunch form factor is clipped at the \SI{10}{\THz} position. In this instance, the bandwidth acts like a low-pass filter, and the bunch length resolution is restricted. Although the characteristics of this particular simulation bandwidth are nonphysical, once real materials are included in the simulation with real transmission bands, this is the type of effect that will be visible. This is the source of the resolution limit mentioned earlier.

\section{Proof of Concept Measurements\label{sec:single_pixel}}
\subsection{Single Pixel Experimental Setup}
\begin{figure*}
\includegraphics[width=1.0\linewidth]{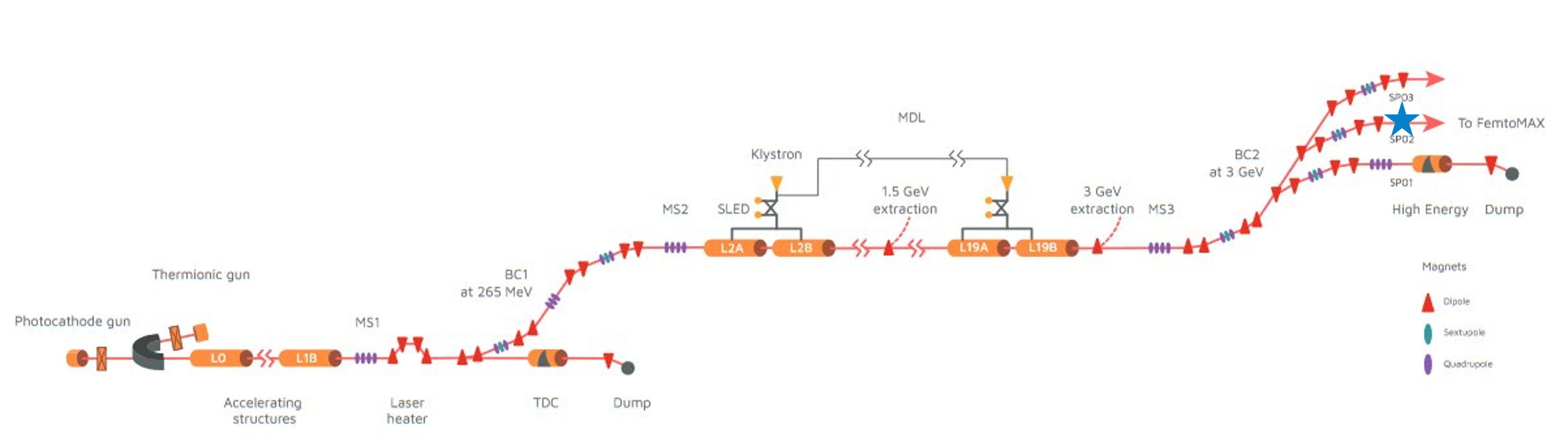}%
\caption{A schematic of the MAX IV linac and the SPF. The location of the CTR imaging setup is indicated with a blue star.\label{fig:maxiv_linac_schematic}}
\end{figure*}

MAX IV is the world’s first operational fourth-generation synchrotron light source~\cite{Thorin2014}. It is composed of a \SI{3}{\GeV} linac feeding two storage rings and a dedicated Short Pulse Facility (SPF)~\cite{Kraljevic2022}. In the SPF, low-emittance electrons are delivered in low-charge, \SI{\sim100}{\pico\coulomb}, for the generation of femtosecond-scale X-ray pulses ideal for time-resolved studies of rapid processes at the atomic scale~\cite{Enquist2018}. Importantly, the MAX IV linac can be tuned to produce a wide range of bunch lengths and longitudinal profiles within the SPF - this made the SPF an ideal location for PoC CTR imaging. A schematic of the MAX IV linac, the SPF, and the position of our PoC setup on the SP02 beamline is presented in Fig.~\ref{fig:maxiv_linac_schematic}.

The longitudinal profile can be varied by two methods in two locations. The first is to vary the phase of the RF pulse, K01 and K02, within either of the two RF accelerating structures, L01 and L02 respectively. It is important to note that all RF stations in a given section, L01 and L02 as shown in Fig.~\ref{fig:maxiv_linac_schematic}, follow the same phase control, i.e. despite there being many RF structures, only two phase values are needed to perform a phase scan, K01 prior to BC1 and K02 prior to BC2. Using K01 or K02 to shift the accelerating gradient experienced by the electron bunch induces an energy chirp, or time-energy correlation. When the bunch then passes through a dispersive beamline, such as either of the two bunch compressor dogleg sections (BC1 and BC2 in Fig.~\ref{fig:maxiv_linac_schematic}), this energy chirp causes longitudinal compression dependent upon the magnitude of the chirp~\cite{Amico2012}. The second method is to alter the strength of the sextupole magnets within the bunch compressor sections. This varies the linearization of the longitudinal phase space, and in doing so, changes the profile shape and compression imparted upon the particle bunch~\cite{England2005}. In these initial measurements the main bunch length controls used were the linac phases, K01 and K02. This was adjusted incrementally to provide a phase scan moving from a region of under-compression, to maximum compression, to over-compression, allowing the broadband CTR image shape and intensity to be monitored across these three regimes.

\begin{figure}
\includegraphics[width=1.0\linewidth]{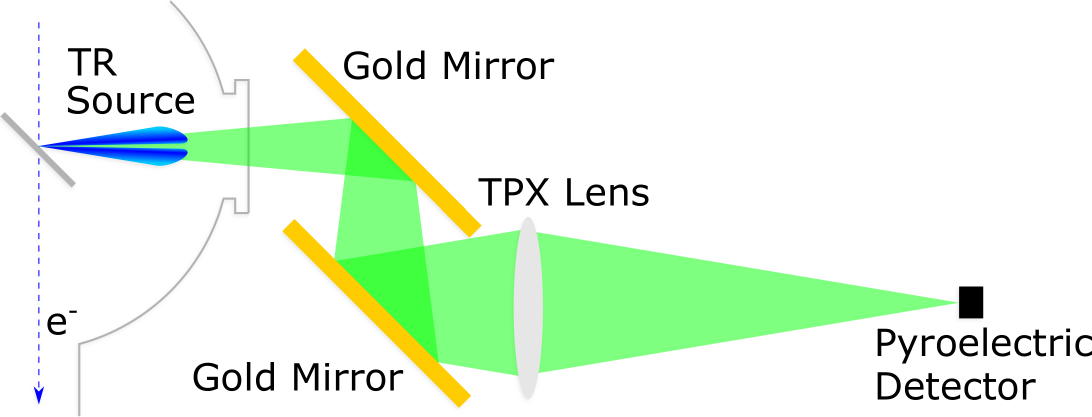}%
\caption{Schematic of CTR single pixel imaging system.\label{fig:single_px_diagram}}
\end{figure}

\begin{figure}
\includegraphics[width=1.0\linewidth]{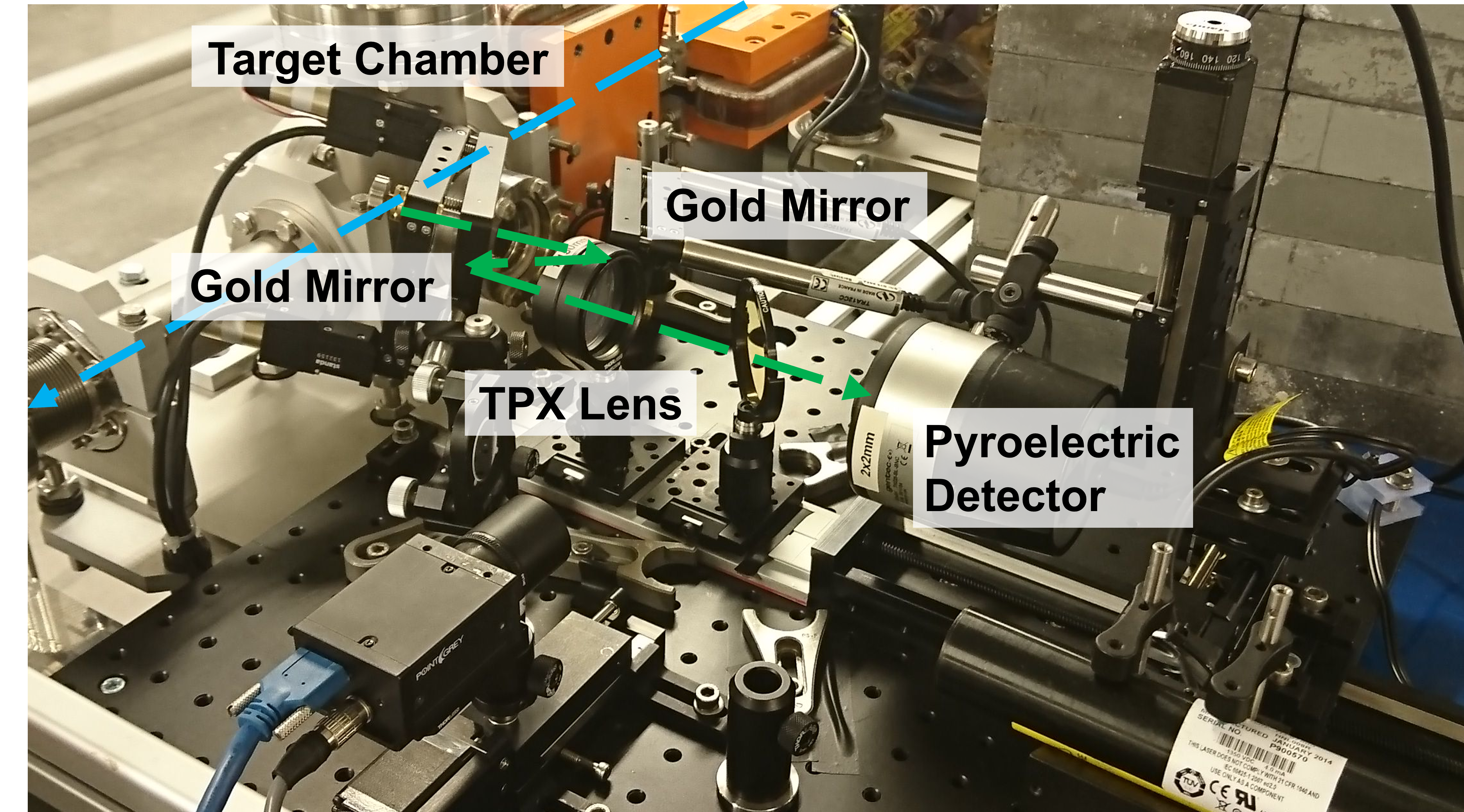}%
\caption{Image of CTR single pixel imaging system with associated alignment and debugging instrumentation.\label{fig:single_px_image}}
\end{figure}

At the time of these measurements, the TDC now in commissioning at MAX IV~\cite{Kraljevic2022} was not available. From previous simulation studies~\cite{Kyle2019}, the bunch lengths of interest were estimated to be $<$\SI{100}{}$-$\SI{300}{\femto\second}. Assuming a Gaussian longitudinal bunch profile as an RMS estimate, the longitudinal form factors for bunch lengths at the extremities coincide with those presented in Fig.~\ref{fig:FF}, and the previously stated bandwidth of \SI{0.1}{\THz} $-$ \SI{10}{\THz}. This information was used to choose the materials for the imaging system. The TR target was a \SI{25}{\mm} diameter aluminium substrate with a \SI{200}{\nm} layer of gold as the incident surface; sufficiently thick for the frequency bandwidth of interest in this case. The viewport window material chosen was Z-Cut Quartz~\cite{Tydex}; this is highly transmissive in the bandwidth of interest and optically, providing the ability to use a simple diode laser for system alignment with the beam path. In these PoC measurements, the imaging system was simplified as much as possible, with only two major components; a single imaging lens and a detector. The lens material chosen was TPX~\cite{Tydex}. A TPX lens provided two benefits, a transmission response highly matched to Z-Cut Quartz, and a refractive index which is approximately constant with varying wavelength. A TPX lens therefore has the same focal length at optical wavelengths as it does in the THz regime, meaning the same diode laser could be used to align and also to focus. A TPX lens with a focal length of \SI{150}{\mm} and a \SI{50}{\mm} diameter was set in a $2f-2f$ configuration to provide $1:1$ imaging of the TR source. The detector was a pyroelectric crystal~\cite{Hook2008}. The exact intensity level for the bandwidth of interest could not be accurately estimated beforehand, so a large format single pixel detector was implemented in a 2D scanning system, as the sensitivity of pyroelectric pixels is proportional to the area of the pixel~\cite{Hook2008}. The resolution of this system was ultimately limited by a combination of the pixel format and the scan resolution, however, the increased pixel size dramatically increased the sensitivity of the detector. The detector chosen (Gentec THZ2I-BL-BNC~\cite{Gentec}) consisted of a \SI{4}{\mm^2} square crystal of Lithium Tantalate (LiTaO3). Additional gold mirrors were placed along the optical path with remotely operable actuators that could adjust horizontal and vertical tilt. These were to be used for online adjustment of the alignment if needed. A complete schematic of the system is presented in Fig.~\ref{fig:single_px_diagram} and an image of the installation in SP02 is presented in Fig.~\ref{fig:single_px_image} along with additional focusing and online debugging optics and equipment.

The detector was placed upon a 3D translation stage. The longitudinal axis was used to adjust the imaging focus, whilst the transverse 2D plane was used to construct a 2D image. For a 2D scan, the actuators set the detector to a particular transverse position within the data collection region, specified by $(x,y)$ coordinates. The detector voltage, averaged over enough shots to reduce the standard deviation below $5\%$, was then assigned as a pixel intensity for that $(x,y)$ image plane coordinate. The scan then moved to the next coordinates and repeated the process, slowly rasterizing a complete image. This functionality led to efficient and scalable scans, as for weaker signals at lower compression, additional measurements at each position were automatically taken to produce the required accuracy. Whereas, for higher compression, less measurements were needed, leading to shorter scans. Thus, all pixel values within all images had the same level of uncertainty.

\subsection{Single Pixel Results}
\begin{figure}
     \centering
     \begin{subfigure}[b]{0.45\textwidth}
         \centering
         \includegraphics[width=\textwidth]{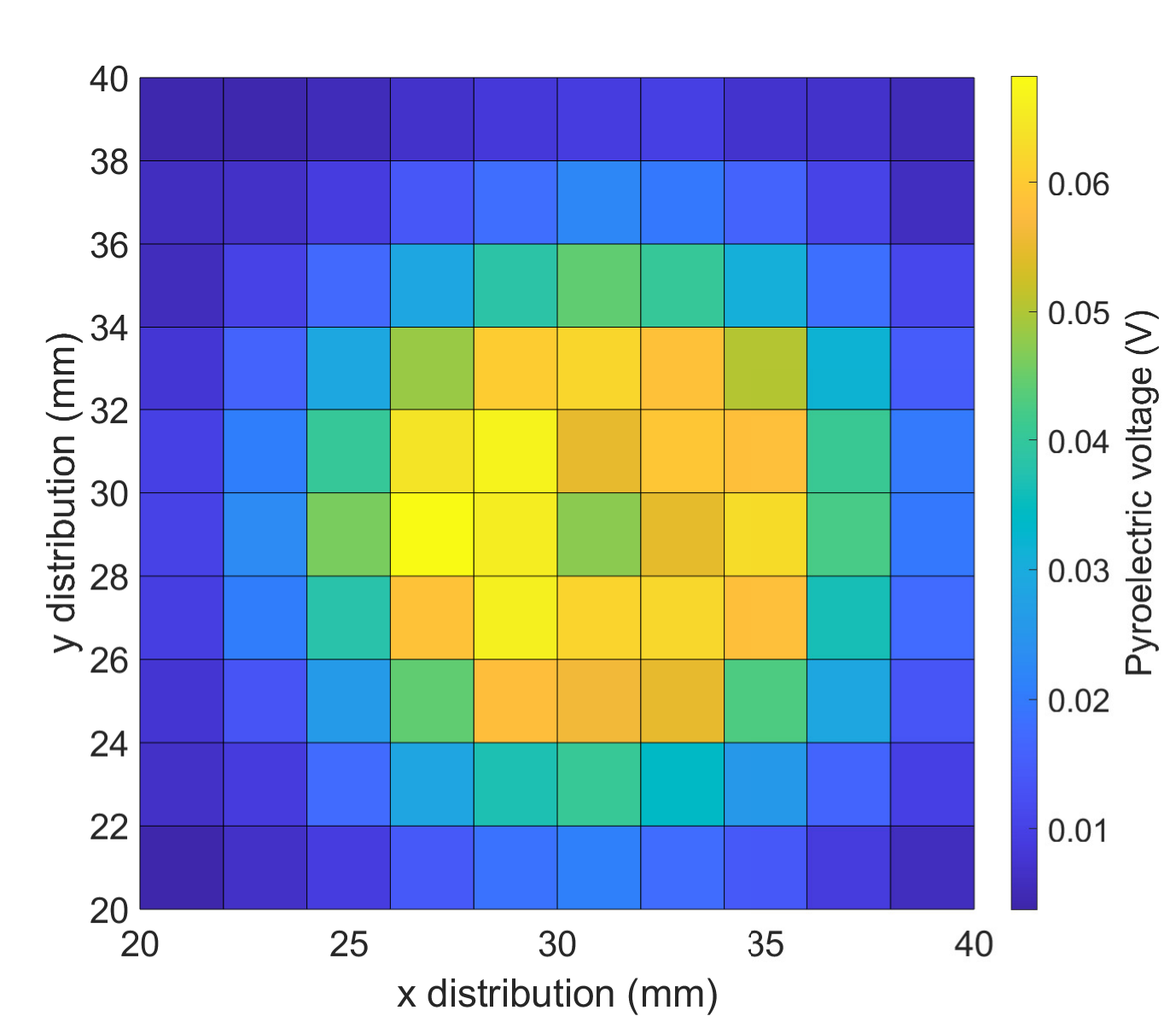}
         \caption{Full radial polarization}
         \label{fig:tpx_polarisation_rad}
     \end{subfigure}
     \vfill
     \begin{subfigure}[b]{0.45\textwidth}
         \centering
         \includegraphics[width=\textwidth]{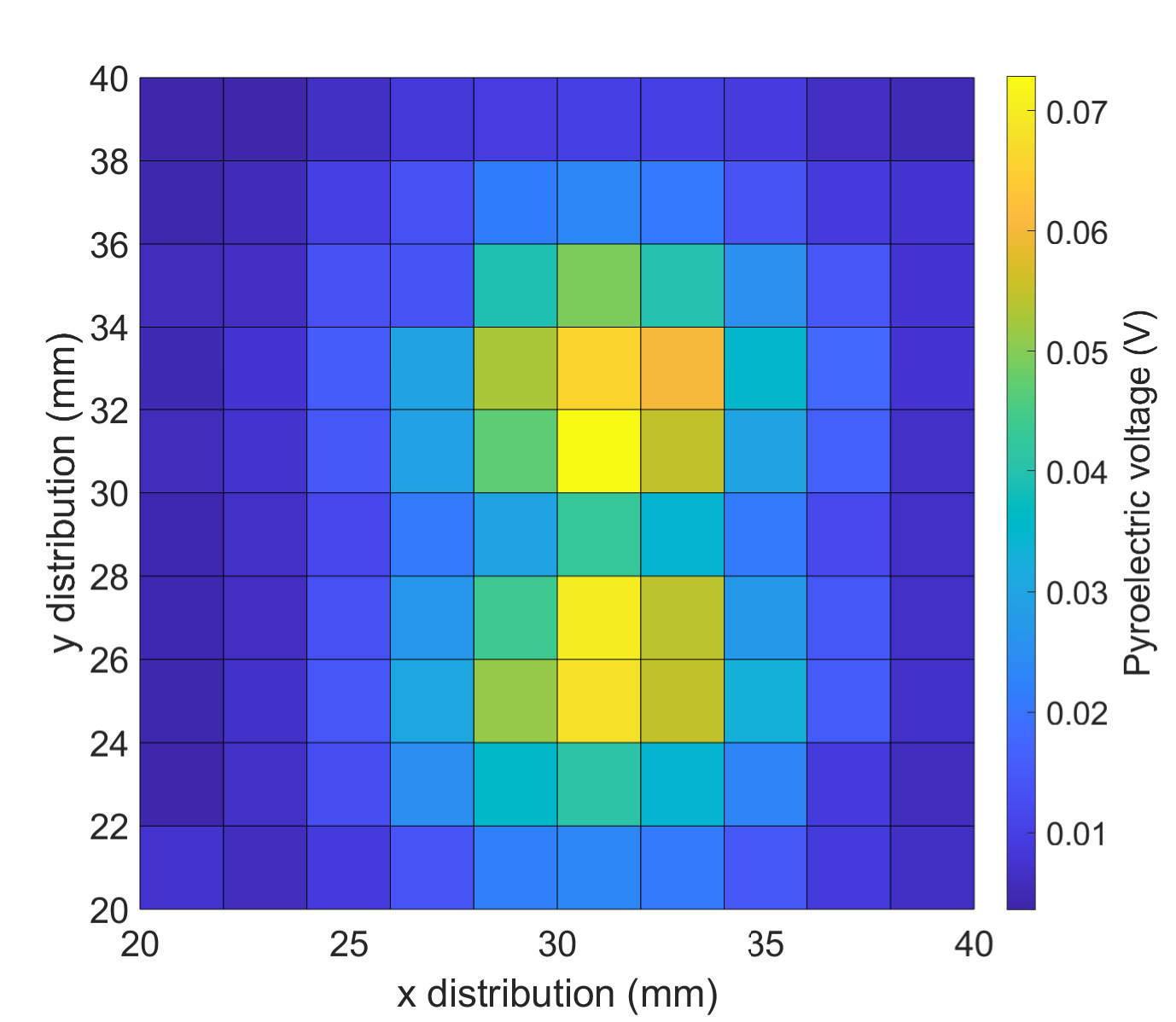}
         \caption{Vertical polarization}
         \label{fig:tpx_polarisation_ver}
     \end{subfigure}
     \vfill
     \begin{subfigure}[b]{0.45\textwidth}
         \centering
         \includegraphics[width=\textwidth]{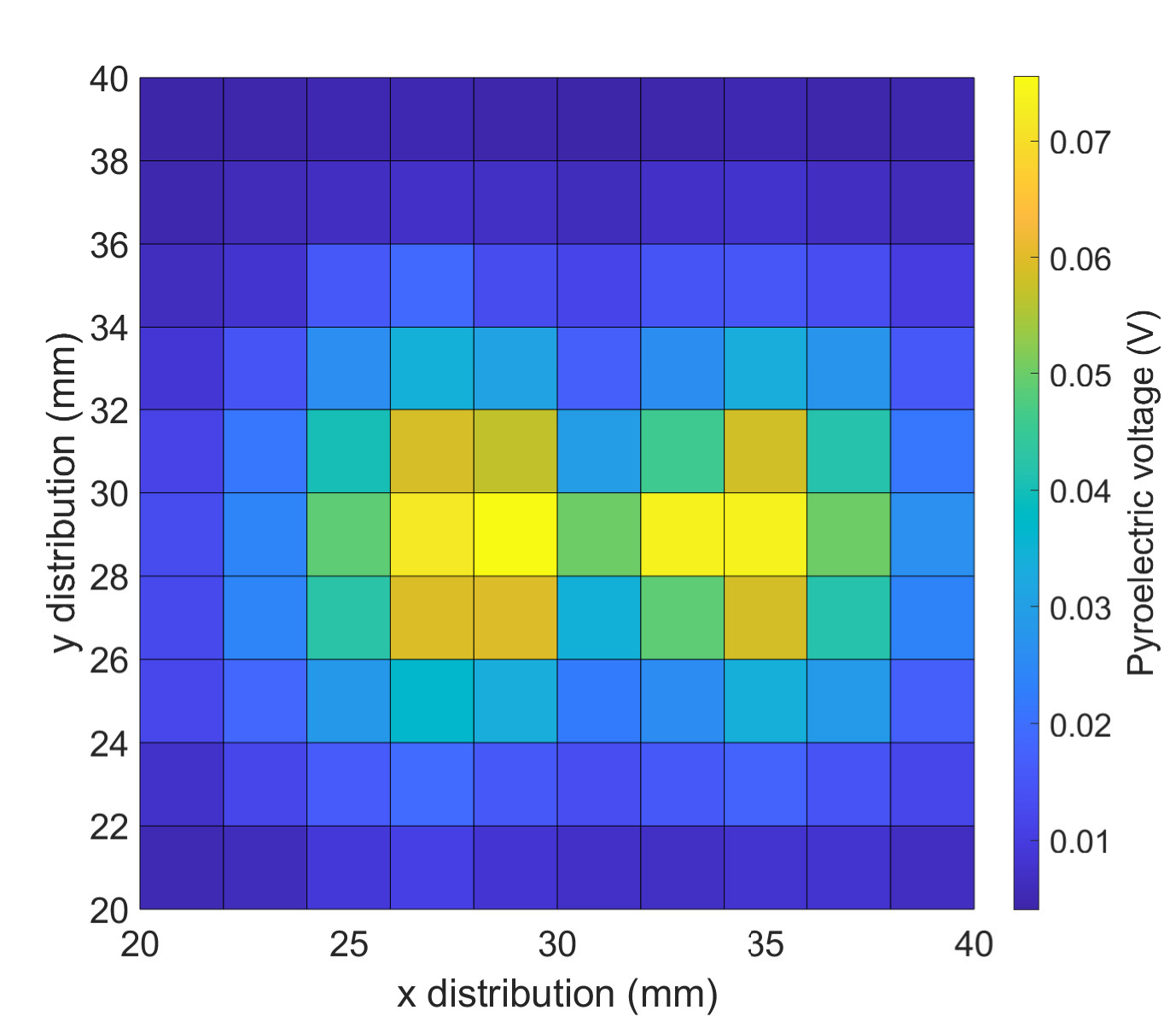}
         \caption{Horizontal polarization}
         \label{fig:tpx_polarisation_hor}
     \end{subfigure}
        \caption{Comparison of measured CTR images using different HDPE polarizer positions.}
        \label{fig:tpx_polarisation}
\end{figure}

To provide definitive evidence that the signal being measured was indeed a CTR source, a high density polyethylene (HDPE) polarizer was used to test for the characteristic radial polarization of the image~\cite{Ter-Mikaelian1972}. Figure~\ref{fig:tpx_polarisation} presents a comparison of radially polarized (no polarizer), vertically polarized, and horizontally polarized CTR images for a fixed compression. As expected, in Fig.~\ref{fig:tpx_polarisation_ver} and Fig.~\ref{fig:tpx_polarisation_hor}, the horizontal and vertical "lobes", respectively, of the annular distribution in Fig.~\ref{fig:tpx_polarisation_rad} have been attenuated. Furthermore, whereas the total intensity has dropped, the peak intensity of the image has been preserved. Therefore, the CTR spatial image distributions compare extremely well with that expected from theory, Fig.~\ref{fig:tpx_simulation}, and with previous measurements at optical wavelengths~\cite{Karataev2011}. These results, coupled with the fact that the only other source of radiation in the vicinity of the TR target is upstream SR which would be de-focused and reduced to a low-intensity background by the point-to-point spatial imaging system, confirms that the signal being measured is indeed spatial CTR.

\begin{figure}
     \centering
     \begin{subfigure}[b]{0.45\textwidth}
         \centering
         \includegraphics[width=\textwidth]{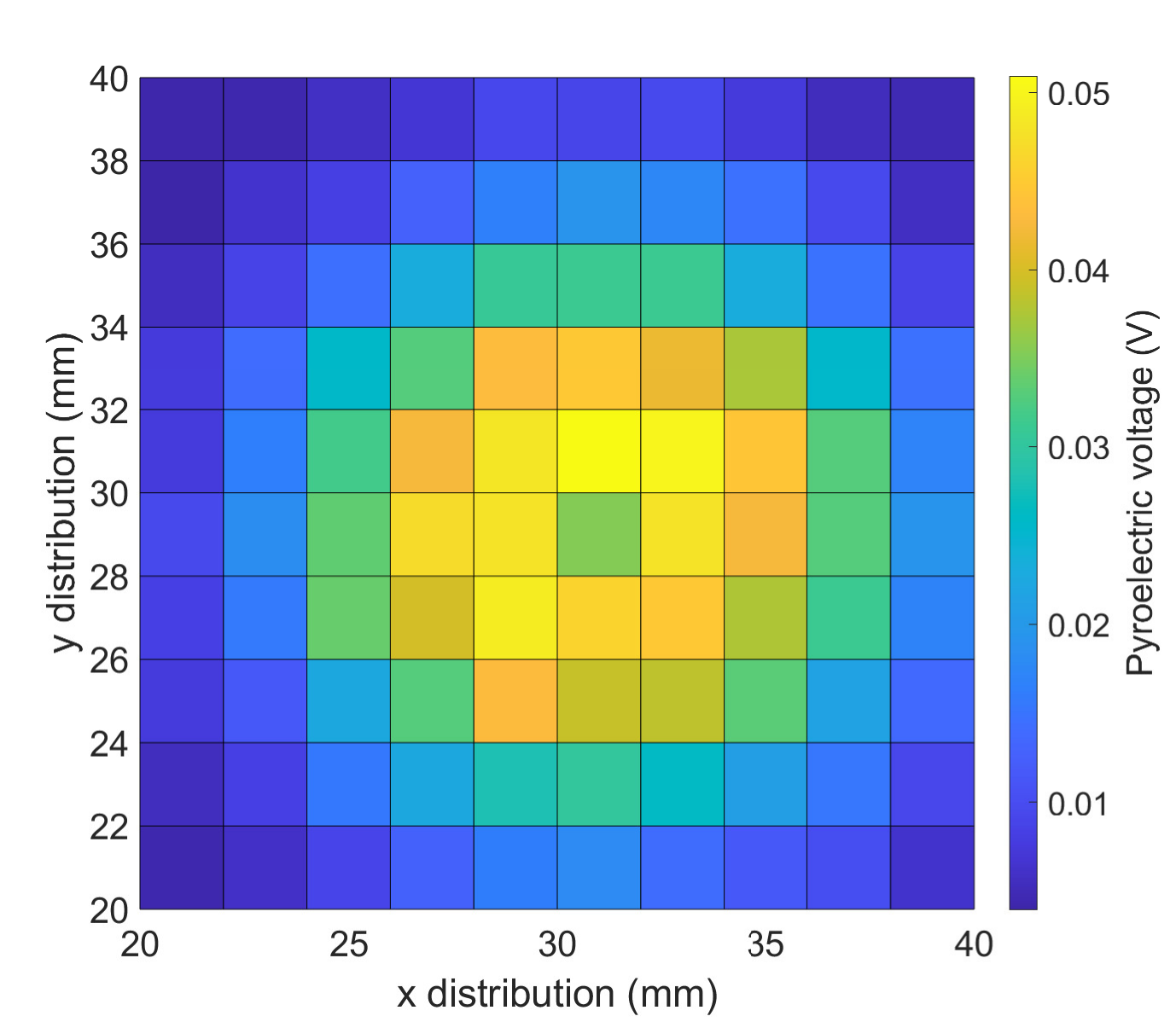}
         \caption{Under-compressed}
         \label{fig:sp_compression_under}
     \end{subfigure}
     \vfill
     \begin{subfigure}[b]{0.45\textwidth}
         \centering
         \includegraphics[width=\textwidth]{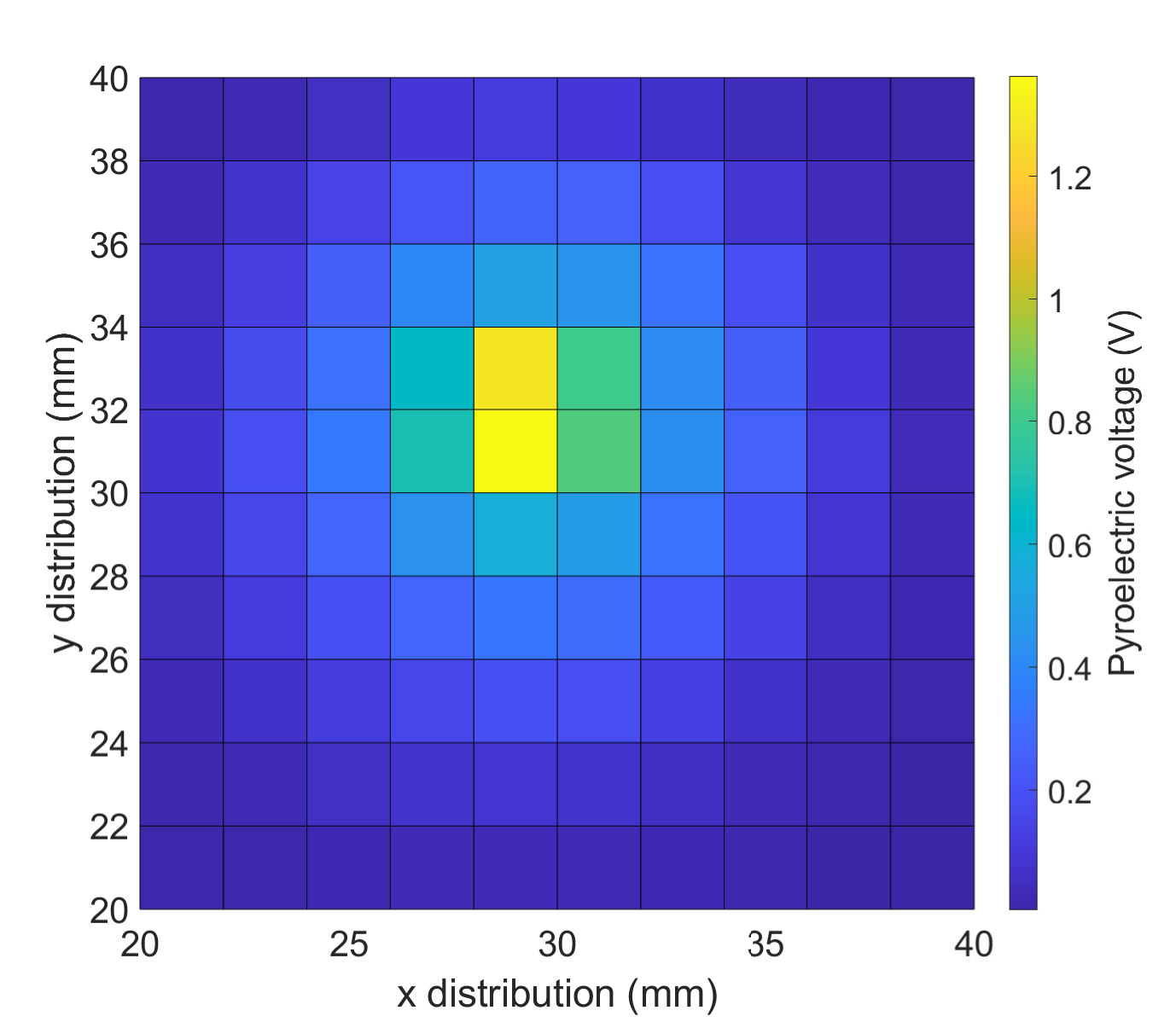}
         \caption{Maximum compression}
         \label{fig:sp_compression_max}
     \end{subfigure}
     \vfill
     \begin{subfigure}[b]{0.45\textwidth}
         \centering
         \includegraphics[width=\textwidth]{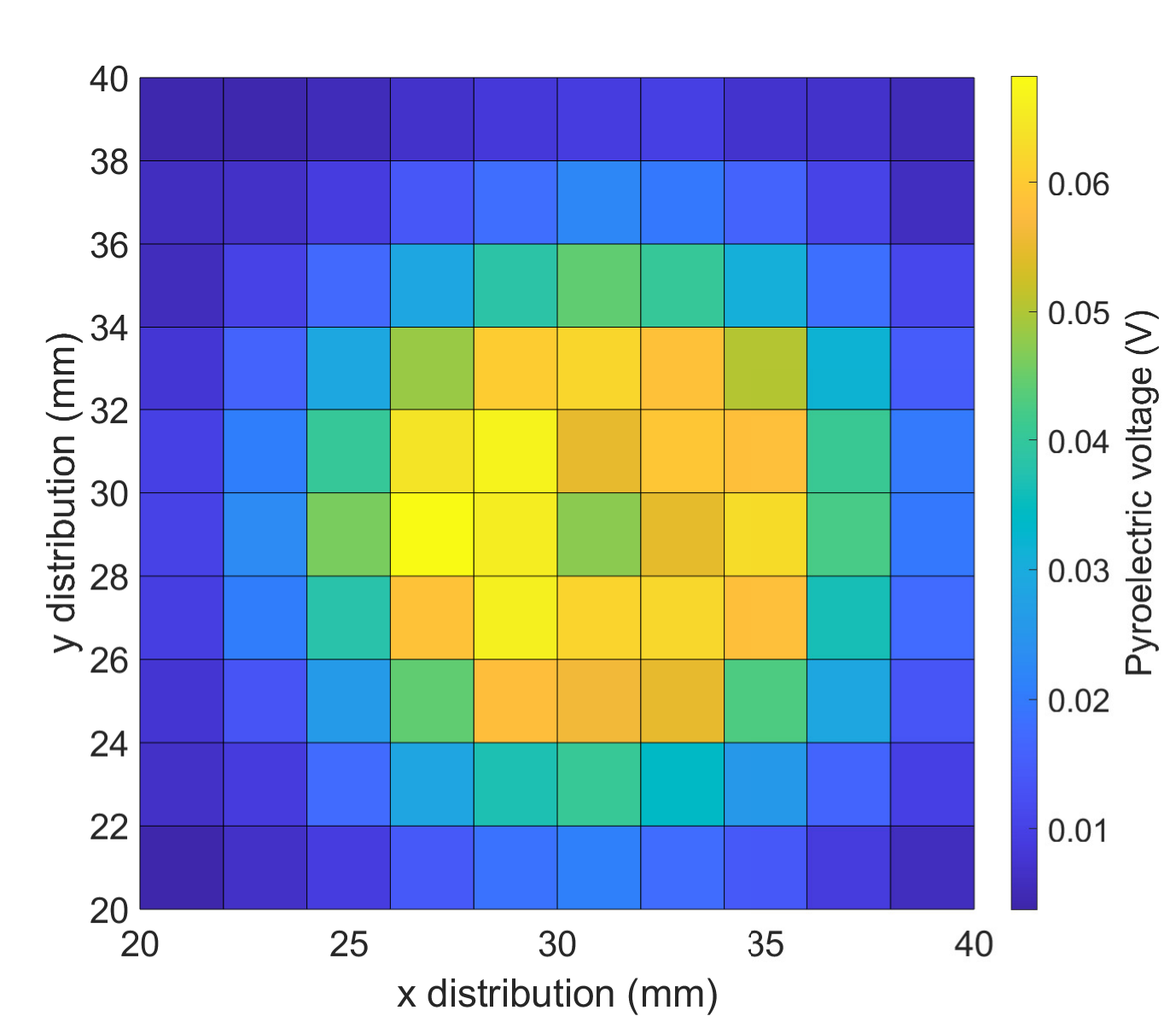}
         \caption{Over-compressed}
         \label{fig:sp_compression_over}
     \end{subfigure}
        \caption{Comparison of measured CTR images at different compressions.}
        \label{fig:sp_compression}
\end{figure}

Following this confirmation, phase scans were taken. The time taken for each raster image to be collected was rather restrictive, therefore, only a selection of positions in the phase parameter space could be measured. These positions were chosen using Elegant~\cite{Borland2000} simulations as a guide, with an emphasis on demonstrating a significant shift in compression. Figure~\ref{fig:sp_compression} presents three examples of the acquired CTR images. For Fig.~\ref{fig:sp_compression_under} and Fig.~\ref{fig:sp_compression_over}, phases were chosen to produce approximately similar bunch lengths, and so similar compressions, but either side of maximum compression - or minimum bunch length. This is evident in the results, as the shape and intensity of the distributions are approximately the same. The characteristic annular shape of the TR signal is also clearly visible, even at this very low resolution, indicating a relatively large bunch length. The \SI{2}{\mm} resolution from the pixel size prevents further analysis of the bunch length with any significant confidence interval. For example, each of the distributions presented in Fig.~\ref{fig:tpx_simulation} would fit inside a single pixel position. This is the case for Fig.~\ref{fig:sp_compression_max}. The image is now so narrow that only a couple of pixel positions dominate the image. Also note the intensity of the image, as this has increased by almost two orders of magnitude. These two effects are exactly as predicted by theory. From Elegant simulations, these phase settings should have produced $\sim$\SI{1000}{\femto\second}, $\sim$\SI{100}{\femto\second}, and $\sim$\SI{1000}{\femto\second}, respectively, where all values are FWHM. These PoC results are in agreement with these estimates.

\begin{figure}
\includegraphics[width=1.0\linewidth]{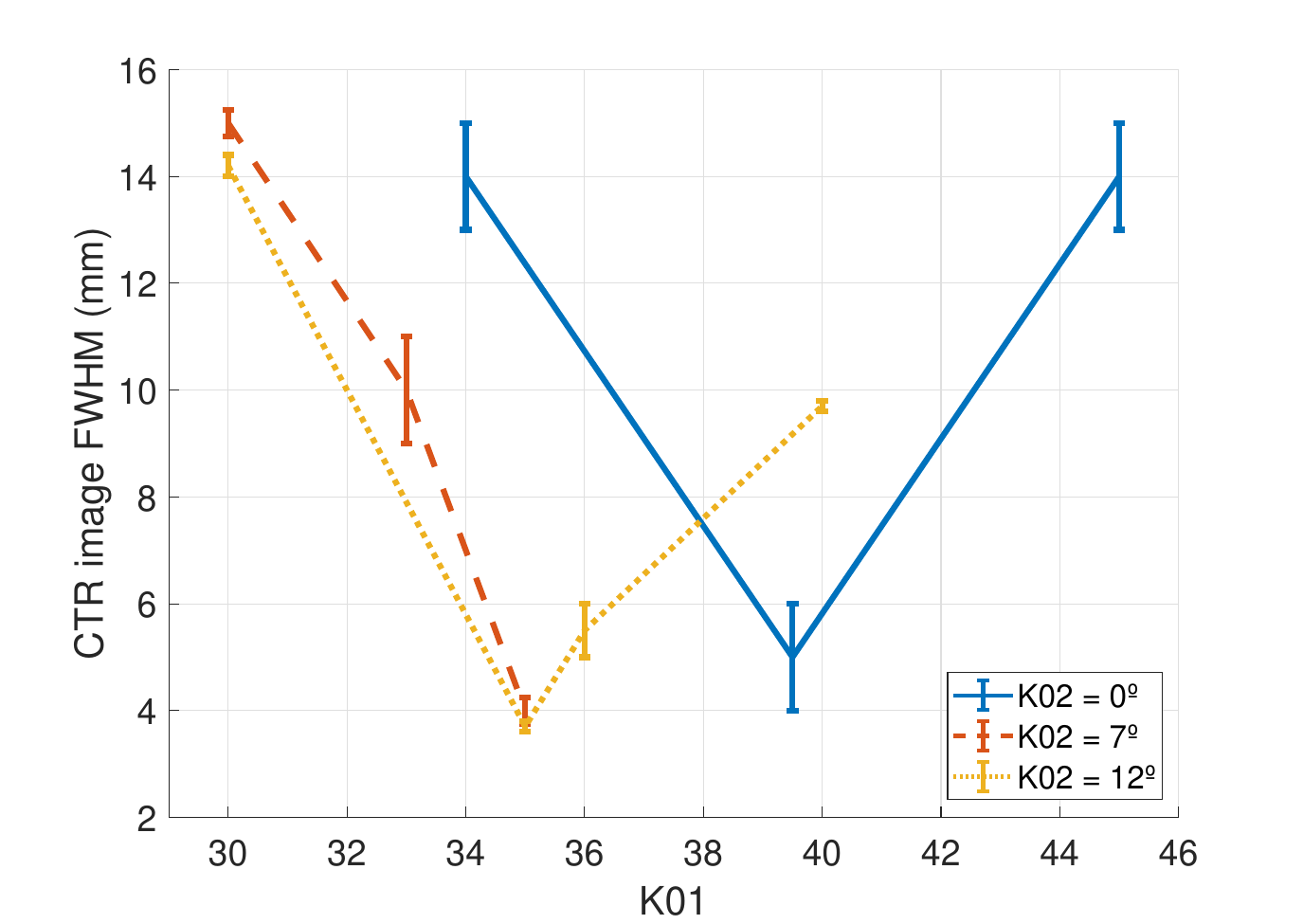}%
\caption{CTR image FWHM variation with  K01 and K02 phase.\label{fig:sp_fwhm_vs_phase}}
\end{figure}

The control of two phase parameters enabled the compression to be probed in different directions, i.e. compressing less in BC1 and more in BC2, and then vice versa. Practically, this was achieved by fixing the K02 phase and then choosing K01 points to image. This was repeated for three different K02 set points. The variation in FWHM of the resulting CTR images is presented in Fig.~\ref{fig:sp_fwhm_vs_phase}. The error bars on the data points have been defined from a combination of the step size and the pyroelectric pixel dimensions. In certain cases smaller steps were taken to test the impact on the resulting image, but very little benefit was found at the expense of a significant increase in scan raster time. 

Unfortunately, the lack of a TDC during these measurements means that the specific bunch lengths at each phase setting cannot be known with any level of certainty. Additionally, comparing with simulation is also challenging due to the significant resolution restriction of the \SI{2}{\mm} wide pixel. However, despite these drawbacks, these measurements clearly demonstrate PoC for both the dependence of the CTR image shape and intensity on longitudinal bunch compression.

There were two main challenges to address to take this PoC and produce an operational piece of instrumentation. First, the excessive time taken to produce each image, and second, the limited spatial resolution in the images. Both of these could be improved by moving away from a single-pixel detector and using an array of smaller pyroelectric pixels.
\section{Compression Monitoring\label{sec:linear_array}}
\subsection{Linear Array Experimental Setup}
\begin{figure}
\includegraphics[width=1.0\linewidth]{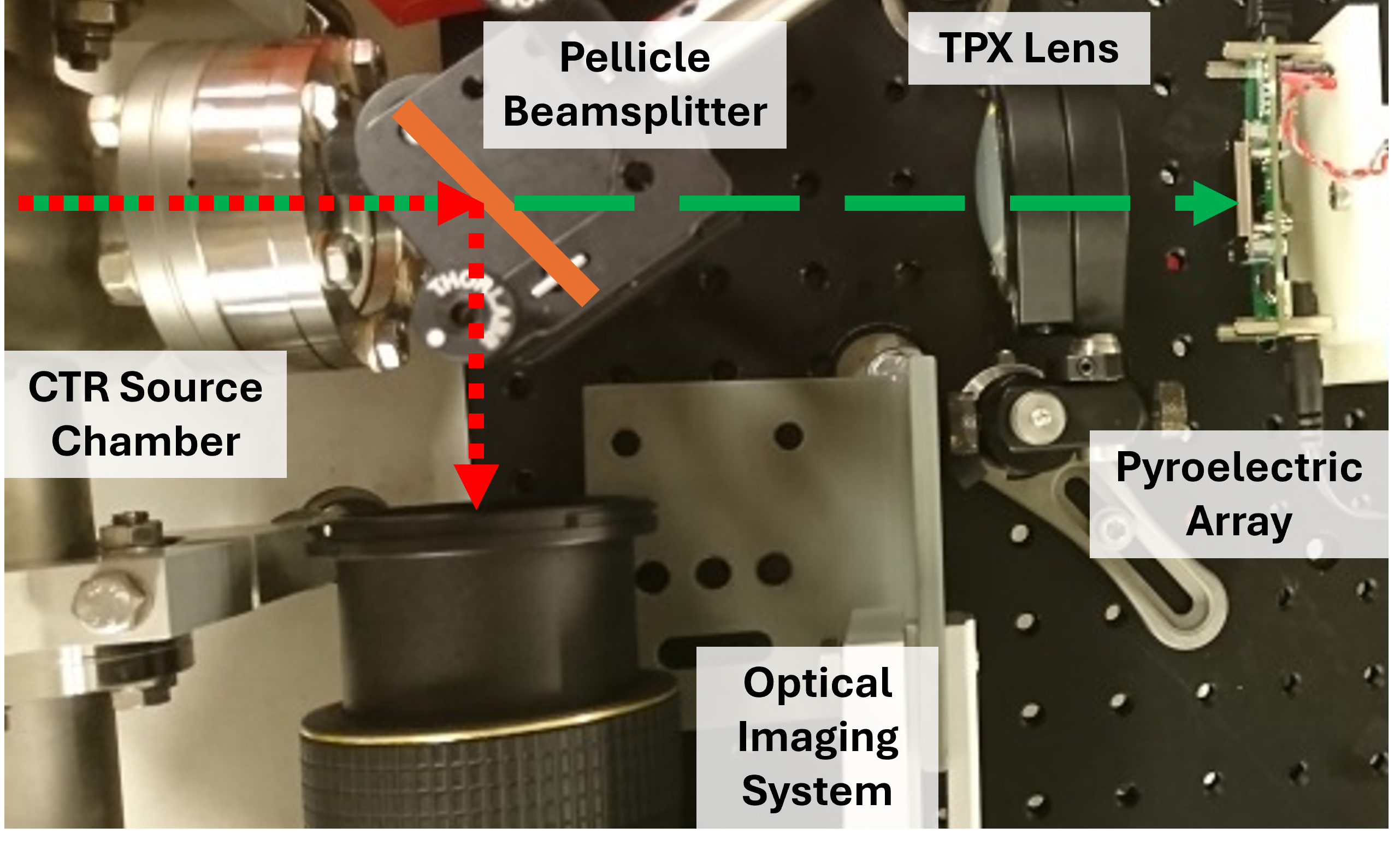}%
\caption{CTR imaging system using linear array of pyroelectric crystals. Both the broadband (green) and optical (red) paths are indicated. The pellicle beam splitter is shown schematically (orange).\label{fig:1d_setup}}
\end{figure}

The setup presented in Fig.~\ref{fig:single_px_image} was simplified in two ways. First, the two gold mirrors were removed; as the additional control they provided was found to be redundant, they added unnecessary complexity. Second, the single pixel detector and scanning system was replaced with a 1D linear array of pyroelectric pixels (DIAS Pyrosens~\cite{DIAS}). This was a 1D array of 256 pixels, each \SI{100}{\micro\m}$\times$\SI{42}{\micro\m}, with a pitch of \SI{50}{\micro\m}. Due to the \SI{5}{\micro\m} thickness of the crystal in each pixel, a protective window was required to prevent damage from optical radiation. High resistivity float zone silicon (HRFZ-Si) was chosen for this purpose due to its extremely flat transmission curve over the bandwidth of interest~\cite{Tydex}. A potential issue with this choice is the associated drop in sensitivity due to the smaller form factor of the pixels and further attenuation by the HRFZ-Si window. Due to these factors, it was expected that this system would be less sensitive to longer bunch lengths but provide finer resolution at shorter bunch lengths. An annotated image of the setup is presented in Fig.~\ref{fig:1d_setup}, where a separate optical beam path is visible in red. The optical light was isolated from the broadband TR using a pellicle beam splitter; not visible in Fig.~\ref{fig:1d_setup} but indicated in orange. The THz signal was unperturbed by the beam splitter due to its thickness of \SI{2}{\micro\m}~\cite{Thorlabs}, therefore, both transverse (optical) and longitudinal (broadband) signals could be acquired from the same TR target. This optical signal was used to align and focus the beam with great effect, ready for the CTR signals to be acquired. 

The linear array was mounted horizontally on a 2-axis motorized stage. The sensor could then be scanned vertically, to locate the CTR image, and longitudinally (along the optical axis), to focus the CTR image. In practice, the vertical stage was adjusted to select the line profile with the lowest central value/maximum visibility between the peaks of the characteristic CTR annular distribution. This ensured the linear array was centered on the CTR image distribution. The longitudinal travel allowed the sensor to be placed in the image plane of the TPX lens and the focal plane. This focal plane capability was not needed for the CTR imaging, but shall be discussed later.

The measurements taken were broadly similar to those using the single-pixel detector, but now these could be captured on a bunch-by-bunch basis. This enabled full phase scans to be completed. Initial results demonstrated that the system was indeed much less sensitive than the single-pixel system. As such, the focusing of the TPX lens was adjusted to de-magnify the CTR source by 3. This increased the signal intensity by a factor of 9, but reduced the spatial resolution across the CTR image. 
\subsection{Linear Array Results}
\begin{figure}
     \centering
     \begin{subfigure}[b]{0.5\textwidth}
         \centering
         \includegraphics[width=\textwidth]{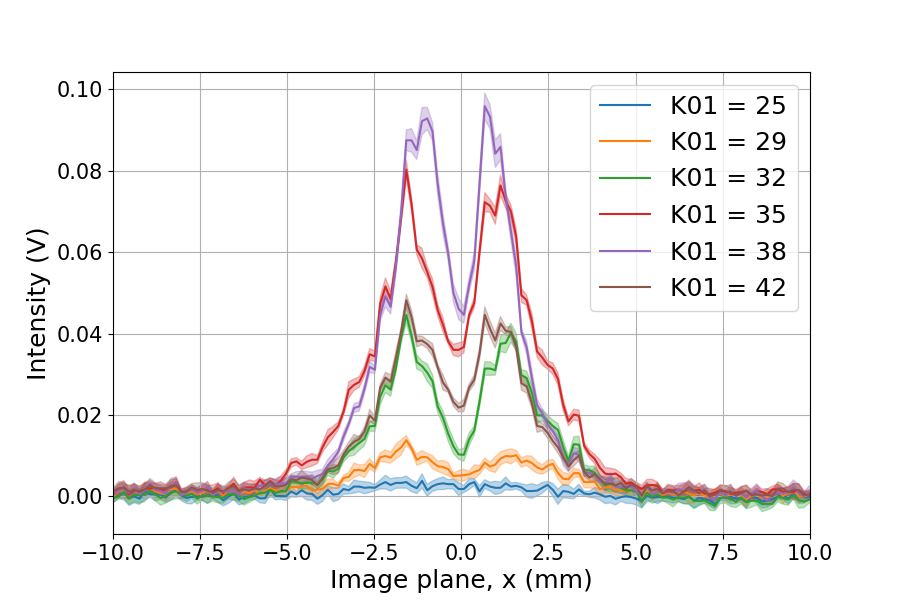}
         \caption{}
         \label{fig:1d_image_comp_abs}
     \end{subfigure}
     \vfill
     \begin{subfigure}[b]{0.5\textwidth}
         \centering
         \includegraphics[width=\textwidth]{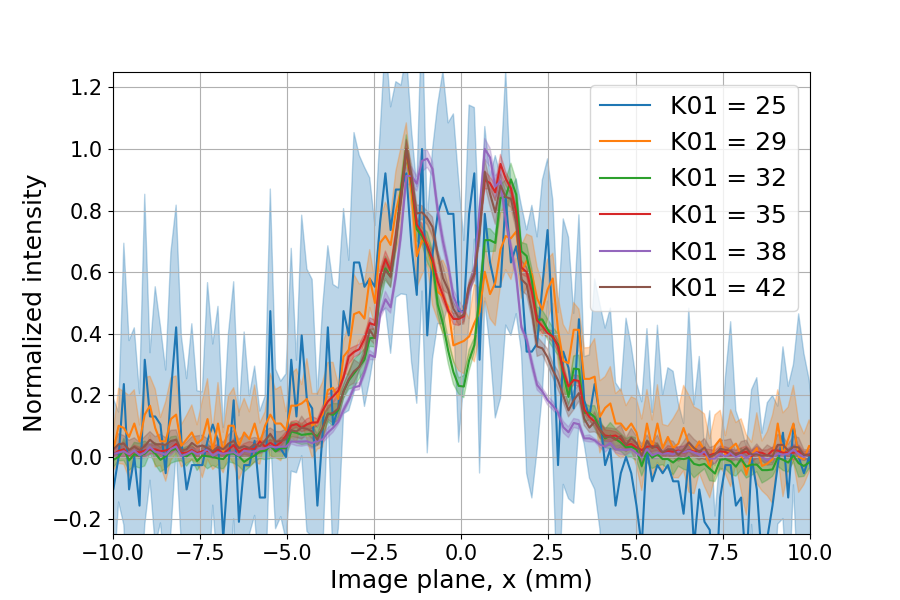}
         \caption{}
         \label{fig:1d_image_comp_norm}
     \end{subfigure}
        \caption{Comparison of measured CTR images at different phases. (a) is the absolute intensity of the images,(b) is the same data normalized to unity for spatial comparison. The shaded area indicates the $\pm\sigma$ uncertainty for the mean of 10 bunches.}
        \label{fig:1d_image_comp}
\end{figure}

Figure~\ref{fig:1d_image_comp} presents CTR line profiles collected within one of the phase scans. K02 was fixed at one of three values, whilst K01 was varied. Although the system appeared to function very well on a bunch-by-bunch basis, the mean and standard deviation of 10 repeat measurements are shown to demonstrate the stability and repeatability of the image profile.

\begin{figure}
     \centering
     \begin{subfigure}[b]{0.5\textwidth}
         \centering
         \includegraphics[width=\textwidth]{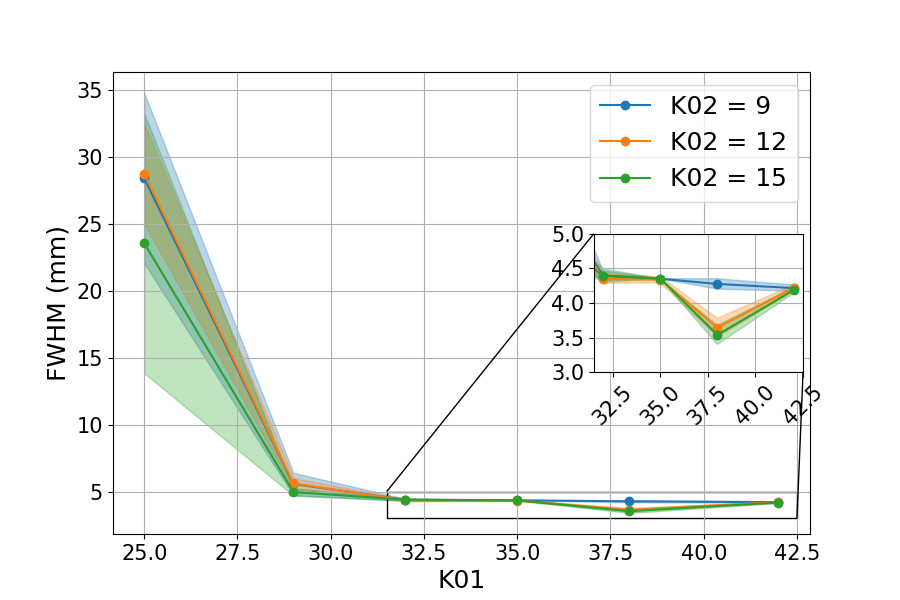}
         \caption{}
         \label{fig:1d_fwhm_variation}
     \end{subfigure}
     \vfill
     \begin{subfigure}[b]{0.5\textwidth}
         \centering
         \includegraphics[width=\textwidth]{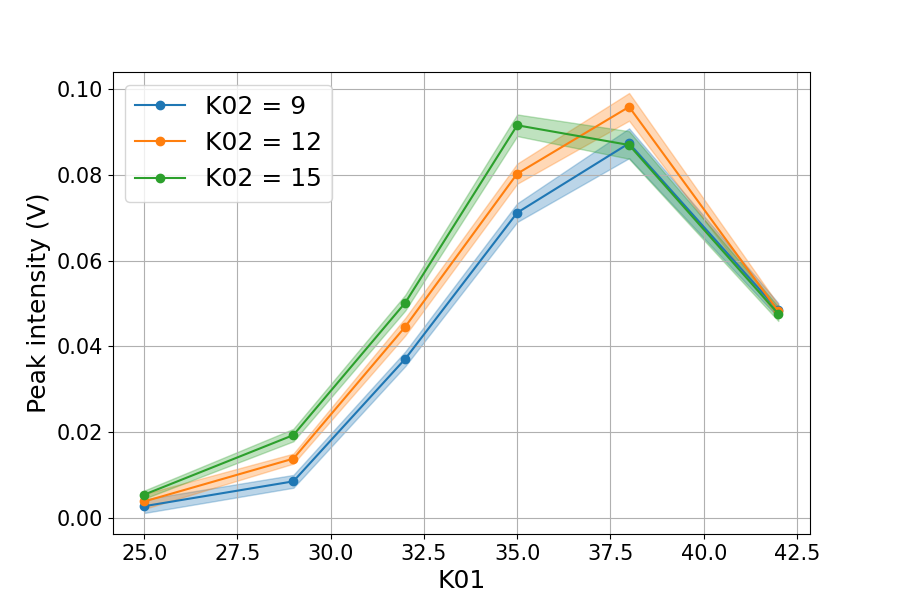}
         \caption{}
         \label{fig:1d_peak_variation}
     \end{subfigure}
        \caption{Variation of CTR image with different phase settings for (a) FWHM, and (b) peak intensity. The shaded area indicates the $\pm\sigma$ uncertainty for the mean of 10 bunches.\label{fig:1d_phase_variation}}
\end{figure}

In Fig.~\ref{fig:1d_image_comp_abs} the variation of the image intensity as a function of compression is clearly visible. In this instance, the bunch charge was monitored as from Equ.~\ref{eq:coh_intensity}, the signal intensity varies as $N^2$. This was reasonably stable at $\sim$\SI{85}{\pico\coulomb} but was varying at the few percent level; the effect of this would be visible in the final results. Another feature to note is the visibility, or the depth of the central minimum of the distributions, which drops, i.e. the visibility becomes worse as the compression increases. This is not predicted by theory; rather, it is an artifact of the form factor of the sensor pixels - specifically the height, \SI{100}{\micro\m}. From the x-axis in Fig.~\ref{fig:1d_image_comp_abs}, it is clear that the CTR image widths are comparable to this pixel height. Therefore, as the image becomes narrower, the CTR image begins to fit vertically upon the array of pixels. This has the effect of filling the central minimum; shorter bunches would eventually saturate this effect. Figure~\ref{fig:1d_image_comp_norm} shows the same scan but normalized. Here, the change in visibility can be seen more clearly, despite the de-magnification described earlier, limiting the extent of the image FWHM variation. An issue with these measurements proved to be the control of the phase and ultimately the compression. As predicted, the system was only sensitive to a narrow range of bunch lengths. If the bunch became too long, the signal intensity dropped before the change in image width could be measured. The phase control used produced non-linear, and quite coarse, changes to the bunch compression. Therefore, a significant increase in the image width is evident when comparing the largest bunch lengths with the shortest, but otherwise the variations are on the few pixel level. This is illustrated in Fig.~\ref{fig:1d_fwhm_variation} which shows how the FWHM of the images in Fig.~\ref{fig:1d_image_comp} vary with phase. Once the magnification of 1/3 has been accounted for, the variation in the zoomed inset region is $\sim$6 pixels on the linear array. Despite this limitation, an obvious minimum region is still clearly visible.

Figure~\ref{fig:1d_peak_variation} shows how the peak intensity varies across each of the phase scans. The peak signal corresponds with the FWHM measurements, which also agrees with the expected position of maximum compression for the MAX IV SPF.

\begin{figure}
     \centering
     \begin{subfigure}[b]{0.5\textwidth}
         \centering
         \includegraphics[width=\textwidth]{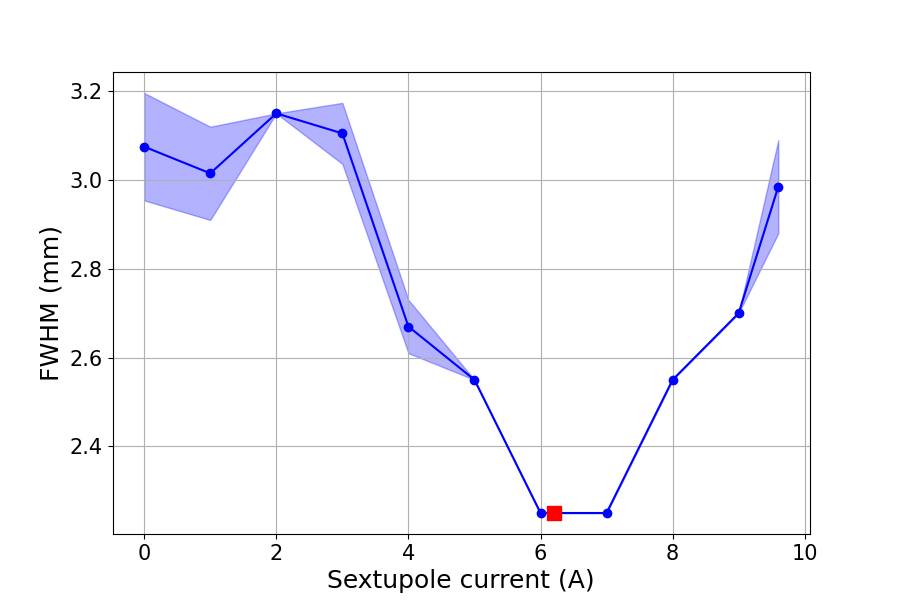}
         \caption{}
         \label{fig:1d_sextupole_fwhm}
     \end{subfigure}
     \vfill
     \begin{subfigure}[b]{0.5\textwidth}
         \centering
         \includegraphics[width=\textwidth]{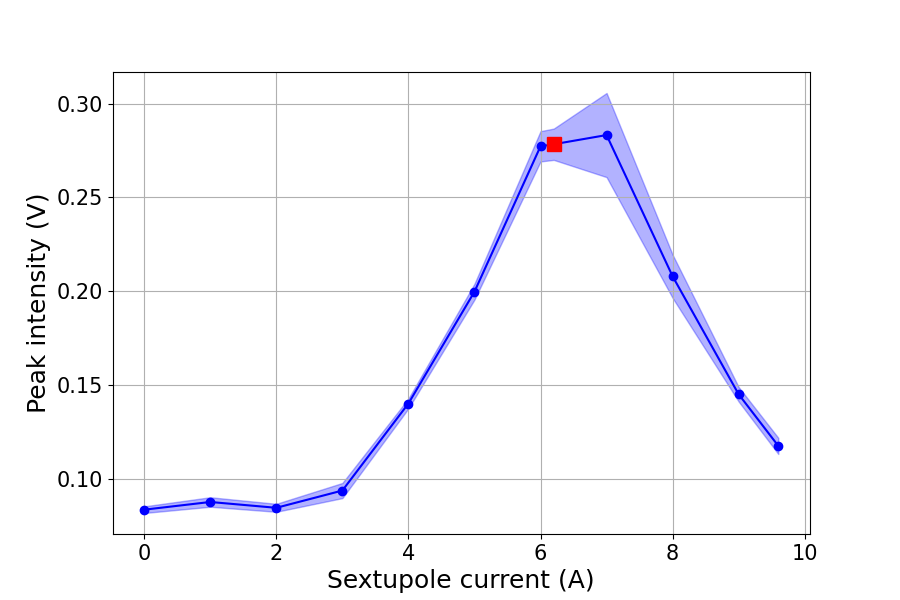}
         \caption{}
         \label{fig:1d_sextupole_peak}
     \end{subfigure}
        \caption{Variation of CTR image across a sextupole scan, for (a) the FWHM, and (b) the peak intensity. The shaded area indicates the $\pm\sigma$ uncertainty for the mean of 10 bunches. The expected optimal value is indicated in red.\label{fig:1d_sextupole}}
\end{figure}

Despite the signal intensity challenges, this broadband imaging method demonstrated self-consistency and agreed with simulations. To produce a smoother compression scan, the linearization of the longitudinal phase space was varied. In dispersive sections, such as BC1 and BC2 in Fig.~\ref{fig:maxiv_linac_schematic}, sextupoles can be used to correct nonlinearities in the LPS by compensating second-order time-energy correlations originating from the RF fields used to chirp the bunch prior to compression~\cite{England2005}. The application of this correction can have a non-linear effect on the longitudinal profile of the bunch, especially so when away from maximum compression. An uncorrected LPS can lead to sharp peaks in the current profile, producing a shorter FWHM bunch length. This happens because the curvature in the LPS can allow particles to redistribute in a way that enhances local charge density at specific longitudinal positions, even as the overall bunch duration increases. In contrast, a fully linear LPS spreads the particles more evenly in time, potentially increasing the measured FWHM. With this process in mind, the sextupole scans were carried out with phase settings adjusted for maximum compression, where in theory maximum compression should be achieved when the LPS is fully linearized; maximizing the CTR image intensity and minimizing CTR image width. The results of one of these scans are presented in Fig.~\ref{fig:1d_sextupole}. Both metrics show a much more gradual variation as planned. A clear minimum in the FWHM is visible in Fig.~\ref{fig:1d_sextupole_fwhm}, which is strongly aligned with the clear peak in intensity in Fig.~\ref{fig:1d_sextupole_peak}. The optimal linearization value expected from simulations is indicated in both figures and is in strong agreement with both measurements.

A point to note is the relatively larger uncertainty in the peak intensity in the region of maximum compression in comparison to the FWHM, where the variation drops to 0 in the same region. Looking at individual image distributions, it was found that the FWHM was very stable, whereas the peak intensity varied shot-to-shot by as much as \SI{20}{\percent}. This can be attributed to shot-to-shot noise in the bunch charge. In the worst case, this would correspond to a charge jitter of $\sim$\SI{10}{\percent}, but even smaller variations will have a large effect given the $N^2$ dependence in Equ.~\ref{eq:coh_intensity}. This type of fluctuation is not unexpected when compression is near maximum or even over-compressed, and serves to highlight the value of the shape/width variation metric.

Once again, the lack of a TDC at the time was a significant limitation to benchmarking this work and converting compression to bunch length. Despite this, a clear dependence of image intensity and width on compression has been demonstrated, with a clear internal consistency. The operational value of such a device has also been demonstrated, with the system working as a fully functional, bunch-by-bunch compression monitor.
\subsection{Towards Non-invasive Operation}
\begin{figure}
     \centering
     \begin{subfigure}[b]{0.5\textwidth}
         \centering
         \includegraphics[width=\textwidth]{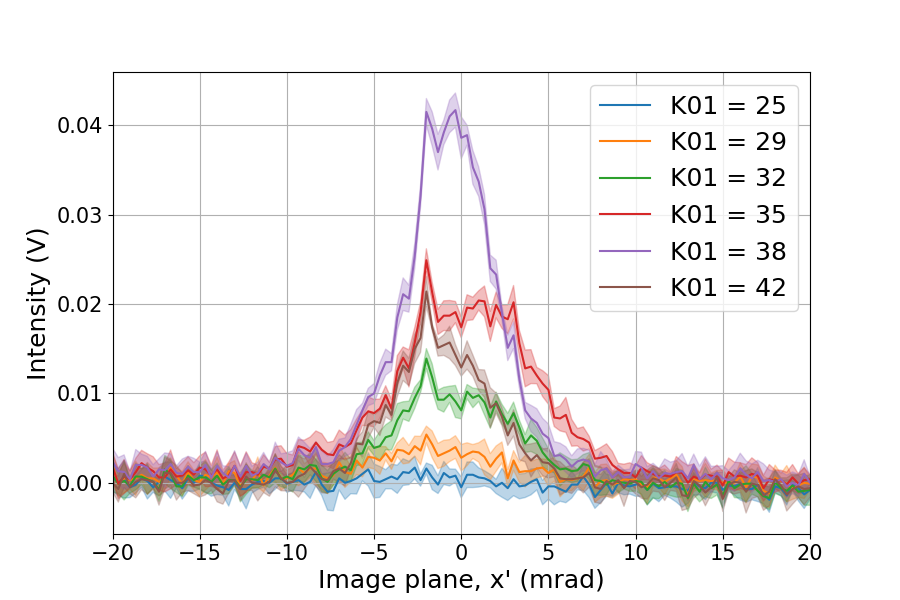}
         \caption{}
         \label{fig:1d_csr_images_phases}
     \end{subfigure}
     \vfill
     \begin{subfigure}[b]{0.5\textwidth}
         \centering
         \includegraphics[width=\textwidth]{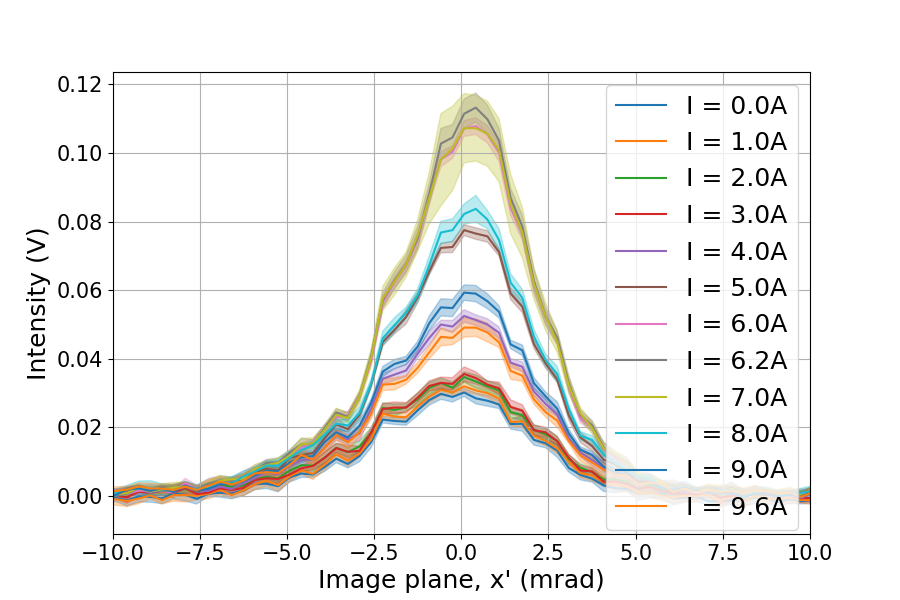}
         \caption{}
         \label{fig:1d_csr_images_sextupole}
     \end{subfigure}
        \caption{Comparison of measured CSR focal plane images for (a) different phases, and (b) a sextupole scan. The shaded area indicates the $\pm\sigma$ uncertainty for the mean of 10 bunches.\label{fig:1d_csr_images}}
\end{figure}

The position of the TR mirror on the MAX IV linac, and the previously described longitudinal translation stage, provided an opportunity to demonstrate a fully non-invasive mode of operation for this device using CSR. This was naturally produced in BC2 rather than through direct interaction with the beam, as for CTR. The exit of the final dipole of BC2 was positioned $\sim$\SI{1}{\m} upstream of the TR mirror. This was too far away to image with the installed setup, however, the translation stage allows the linear array to be positioned in the focal plane of the TPX lens. This allows for far-field angular imaging of the CSR source within the dipole. The only other adjustment to the setup was to partially extract the TR mirror from the beamline so that the beam no longer directly interacted with it. This has the effect of changing the CTR source mirror into a CSR extraction half-mirror. With this slight modification, the exact same scans could be performed, but this time using non-invasive CSR. Figure~\ref{fig:1d_csr_images} presents the CSR results for the same phase scan as in Fig.~\ref{fig:1d_image_comp} and the sextupole scan in Fig.~\ref{fig:1d_sextupole}.

Despite the obvious difference in the overall shape of the CSR distribution in comparison to the CTR distribution, the trends in the data are exactly the same. A small difference is that the overall intensity is lower in the CSR images compared to the CTR images. This difference could be attributed to the detector being placed in the focal plane of the TPX lens rather than in the image plane, as in this instance the image distribution is narrower than the angular distribution.

\begin{figure}
     \centering
     \begin{subfigure}[b]{0.5\textwidth}
         \centering
         \includegraphics[width=\textwidth]{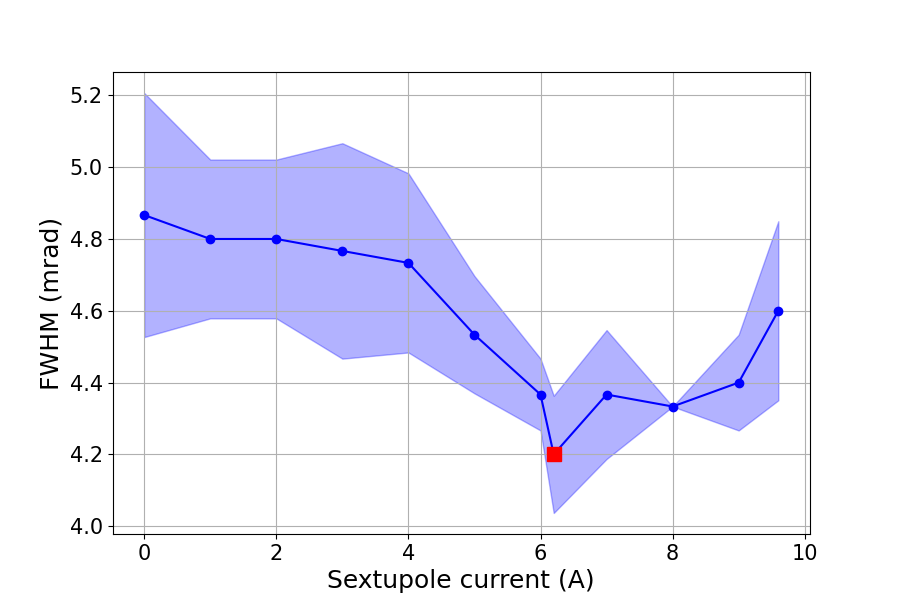}
         \caption{}
         \label{fig:1d_csr_sextupole_fwhm}
     \end{subfigure}
     \vfill
     \begin{subfigure}[b]{0.5\textwidth}
         \centering
         \includegraphics[width=\textwidth]{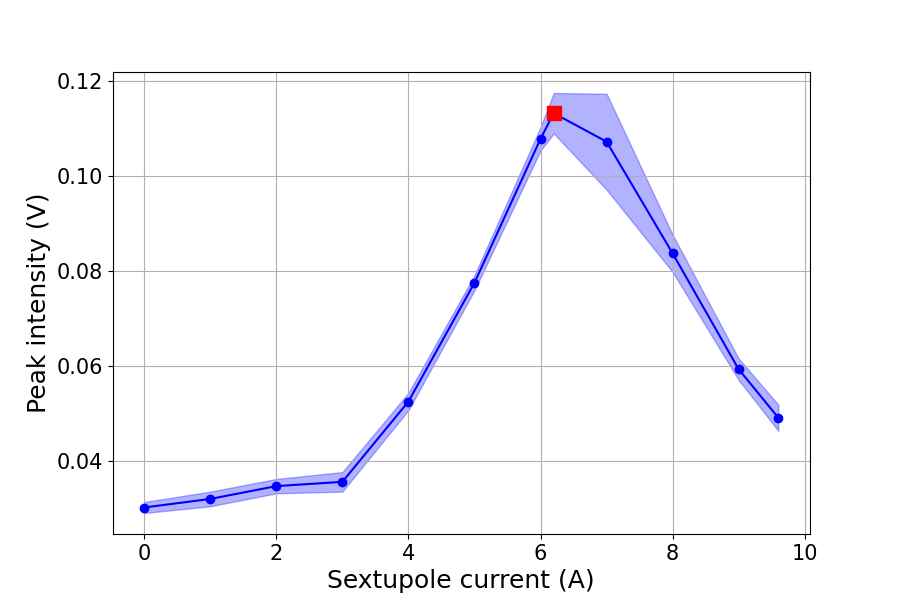}
         \caption{}
         \label{fig:1d_csr_sextupole_peak}
     \end{subfigure}
        \caption{Variation of CSR image across a sextupole scan, for (a) the FWHM, and (b) the peak intensity. The shaded area indicates the $\pm\sigma$ uncertainty for the mean of 10 bunches. The expected optimal value is indicated in red.\label{fig:1d_csr_sextupole}}
\end{figure}

Figure~\ref{fig:1d_csr_sextupole} shows how the FWHM of the image and the peak intensity vary for the sextupole scan presented in Fig.~\ref{fig:1d_csr_images_sextupole}. The results are almost identical to those found in Fig.~\ref{fig:1d_sextupole}, except for a minor relative increase in the uncertainty level, which is directly related to the lower signal-to-noise level. The change in angular width between the largest and smallest FWHM corresponds to $\sim$3 pixels on the linear array, which was expected since these measurements were taken in the focal plane of the imaging lens and were not optimized for point-to-point imaging as with the CTR. Despite this slight variation, the minimum in FWHM and maximum in peak intensity correspond to one another, agree with the CTR results, and match the expected optimal current for longitudinal phase space linearization. These results provide another PoC, but in this instance, for compression monitoring on a bunch-by-bunch basis via non-invasive broadband imaging.

\section{Conclusions\label{sec:conclusion}}
This work has demonstrated PoC results from two methods of monitoring longitudinal bunch compression on a bunch-by-bunch basis. The first utilizes CTR, which provides a relatively strong signal and is simple to implement. The second utilizes non-invasive CSR, providing an online monitoring technique.

These results and theoretical underpinning can be used to apply this technique to any bunch length range. Due to the particularities of polarization-type radiation~\cite{Karlovets2009}, the shorter the bunch length, the easier it is to implement and the stronger the signal becomes. For example, at $\sim$\SI{1}{\femto\s} bunch lengths, the dominant radiation bandwidth is in the infrared, which means infrared optics and cameras can be used, considerably simplifying the system. This could make this technique extremely attractive to novel acceleration schemes such as laser plasma wakefield acceleration~\cite{Walker2017}.

There are two main improvements required for this system. The first is to improve the signal level. The low peak intensity, per unit charge, limits the bunch length range of applicability for this method. A new optical system using reflective optics would minimize losses from transmission curves and could be tuned to produce a much smaller point spread function (PSF). In doing so, the increased signal level would allow an increase in the magnification, and therefore, an increase in the spatial resolution for FWHM measurements. The other improvement would be to link the CTR images to the bunch length. This could be achieved via simulation studies, but would require a TDC to benchmark against. A comparison at this stage would be pure conjecture. As MAX IV finishes commissioning a TDC beamline, these measurements will become a priority.

To enable future non-invasive operation, a dedicated CSR imaging system will be required. While placing the detector in the focal plane was sufficient for PoC purposes, this approach suffers in practice from reduced signal strength and interference from upstream sources; particularly so at longer wavelengths and higher beam energies. Ideally, a system could switch between imaging the source point of both CTR and CSR, but in most cases this is impractical. This underscores the need for a dedicated CSR diagnostic, for which this work has provided an excellent foundation.

This work has, for the first time, demonstrated that both the image shape and intensity of broadband coherent radiation can be used for compression monitoring, paving the way for a future non-invasive bunch-by-bunch longitudinal bunch profile monitor.
\begin{acknowledgments}
The authors would like to thank Dr. Ralph Fiorito for his endless enthusiasm for this work and his drive to push it forward. His presence and expertise are still missed across the beam instrumentation community. The authors would also like to thank the MAX IV operators who ran the beam during the data taking. They ensured operation and data collection ran smoothly. This work is supported by the AWAKE-UK project funded by STFC grant No. ST/R002312/1, the STFC Cockcroft core grant No. ST/G008248/1, and the European Union’s Horizon Europe research and innovation program under grant agreement no. 101073480 and the UKRI guarantee funds.

The data that support the findings of this article are openly available~\cite{WolfendenData2025}.
\end{acknowledgments}

% Create the reference section using BibTeX:
\bibliography{references}

\end{document}